\documentclass[12pt]{article}
\usepackage{amsmath,amsfonts,amssymb}
\newcommand{\be}{\begin{equation}}
\newcommand{\ee}{\end{equation}}
\newcommand{\ba}{\begin{eqnarray}}
\newcommand{\ea}{\end{eqnarray}}

\newcommand{\la}[1]{\label{#1}}

\def\gl#1{(\ref{#1})}

\date{}
\begin{document}
\title{ Non-linear Supersymmetry for non-Hermitian, non-diagonalizable
Hamiltonians: II. Rigorous results}
\author{A. V. Sokolov\\
{ \it  V.A.Fock Institute of Physics, Sankt-Petersburg State University}}
\maketitle
\abstract{
 We continue our investigation of the non-linear SUSY    for complex potentials started in the Part I
 \cite{ACS} and prove
the  theorems characterizing its structure in the case of non-diagonalizable Hamiltonians. This part provides
the mathematical basis of previous studies. The classes of potentials invariant under SUSY transformations for
non-diagonalizable Hamiltonians are specified and the asymptotics of formal eigenfunctions and associated
functions  are derived.  Several results on the normalizability of associated functions at infinities are
rigorously proved. Finally the Index Theorem on relation between Jordan structures of intertwined Hamiltonians
depending of the behavior  of elements of canonical basis of supercharge kernel at infinity is proven. }

\section{Introduction: definitions and notation}
In this part of the paper we continue the investigation of the nonlinear SUSY  \cite{ais}--\cite{ansok} (see,
the extended list of references in \cite{ACS})  for complex potentials started in the Part I \cite{ACS} and
prove the  theorems characterizing its structure in the case of non-diagonalizable Hamiltonians. We use the
class of potentials invariant under SUSY transformations for non-diagonalizable Hamiltonians and prove several
results concerning the normalizability of associated functions at $+\infty$ or $-\infty$. These results allow to
unravel the relation between Jordan cells in SUSY partner Hamiltonians which was described in the Index Theorem
in Sec 6. of \cite{ACS}.

All the proofs and results of this part are safely applicable to PT symmetric non-Hermitian Hamiltonians \cite{ptsym,jones,bender} with
soft type \cite{ACS} of non-Hermiticity when the real part of a potential dominates over its imaginary one at both
coordinate infinities. The latter property is embedded into the chosen classes of potentials.

Let us summarize this part of the paper aimed to derive
 Theorem~3
and  Lemmas~1--4 discussed in  Part~I \cite{ACS}. First we introduce   the relevant classes of potentials $K$
(main) and $\cal K$ (auxiliary) as well as we remind the notion of formal associated functions of a Hamiltonian.
Next we provide  necessary estimates for potentials belonging to the class $\cal K$ (Lemma~5) and for auxiliary
integrals (Lemmas~6,~7). Furthermore, we  derive the asymptotics of formal eigenfunctions (Lemma~8) and
associated functions (Lemma~9) of Hamiltonians with potentials belonging to the class $\cal K$. Then the
invariance of  the classes $\cal K$ and $K$ under intertwining is proved (Lemmas~10 and~1 respectively). And
finally the proofs of the Lemma~2 (on properties of a sequence of formal associated functions under
intertwining), of the Lemma~3 (on normalizability of elements of the canonical basis of an intertwining
operator), of the Lemma~4 (on interrelation in (non)normalizability of canonical bases of mutually transposed
intertwining operators) and of the Theorem~3 (on relation between Jordan structures of intertwined Hamiltonians
depending on the asymptotic behavior  of elements of the canonical basis of an intertwining operator kernel at
$\pm\infty$ ) will be presented. The enumeration of definitions and corollaries in brackets corresponds to the
enumeration of the same definitions and corollaries in Part~I \cite{ACS}.

In the paper we use the following classes of potentials.

{\bf Definition 1 (2).} Let $K$ be the set of all potentials $V(x)$ such that:

1) $V(x)\in C_{\mathbb R}^\infty$;

2) there are $R_0>0$ and $\varepsilon>0$ ($R_0$ and $\varepsilon$ depend on $V(x)$) such that for any
$|x|\geqslant R_0$ the inequality ${\rm{Re}}\,V(x)\geqslant\varepsilon$ takes place;

3) \be{\rm {Im}}\,V(x)/{\rm {Re}}\,V(x)=o(1),\qquad x\to\pm\infty;\ee

4) functions \be \bigg(\int\limits_{\pm R_0}^x\sqrt{|V(x_1)|}dx_1\bigg)^2
\bigg({{|V'(x)|^2}\over{|V(x)|^3}}+{{|V''(x)|}\over{|V(x)|^2}}\bigg) \ee are bounded respectively for
$x\geqslant R_0$ and $x\leqslant -R_0$.

{\bf Definition 2.} Let $\cal K$ be the set of all potentials $V(x)$ such that:

1) $V(x)$ is a complex-valued (in particular, real-valued) function, defined on the real axis with possible
exception at some points;

2) there are $R_0>0$ and $\varepsilon>0$ ($R_0$ and $\varepsilon$ depend on $V(x)$) such that
$V(x)\Big|_{[R_0,+\infty[}\in C^2_{[R_0, +\infty[}$, $V(x)\Big|_{]-\infty,-R_0]}\in C^2_{]-\infty,-R_0]}$ and
for any $|x|\geqslant R_0$ the following inequality holds:
\begin{equation}{\rm {Re}}\,V(x)\geqslant\varepsilon;\la{a2}\end{equation}

3) \begin{equation}{\rm{Im}}\,V(x)/{\rm{Re}}\,V(x)=o(1),\qquad x\to\pm\infty; \la{a3}\end{equation}

4) functions
$$\bigg(\int\limits_{\pm R_0}^x\sqrt{|V(x_1)|}dx_1\bigg)^2\bigg(
{{|V'(x)|^2}\over{|V(x)|^3}}+{{|V''(x)|}\over{|V(x)|^2}}\bigg)$$ are bounded respectively for $x\geqslant R_0$
and $x\leqslant-R_0$.

Let us clarify that $K$ is a main class of potentials --- the class of physical potentials, and $\cal K$ is an
auxiliary, wider class of potentials --- the class, containing potentials of intermediate Hamiltonians
(corresponding to factorization of an intertwining operator in the product of intertwining operators of first
order in derivative) in the case when potentials of the initial and final Hamiltonians belong to $K$.

In what follows we shall use the functions of the form $(V(x)- \lambda)^\varkappa$, where $V(x)\in\cal K$,
$\varkappa>0$, $\lambda\in\mathbb C$ and either $\lambda\leqslant0$ or ${\rm{Im}}\,\lambda \ne0$. Branches of
these functions will be identically selected by the condition
\begin{equation}|\arg(V(x)-\lambda)|<\pi.\la{a1}\end{equation}
In the case $\lambda\leqslant0$ this condition can be fulfilled in view of \gl{a2}, and in the case
${\rm{Im}}\,\lambda \ne0$ because of \gl{a3} the condition \gl{a1} can be satisfied for any $|x|>R_2$, where
$R_2\geqslant R_0$ is such that for any $|x|\geqslant R_2$ the inequality
\be{{|{\rm{Im}}\,V(x)|}\over{|{\rm{Re}}\,V(x)|}}\leqslant{1\over2}{{|{\rm{Im}}
\,\lambda|}\over{|{\rm{Re}}\,\lambda|}},\qquad |x|\geqslant R_2\la{star}\ee holds. For $-R_0<x<R_0$ (if
$\lambda\leqslant0$) or for $-R_2<x<R_2$ (if ${\rm{Im}}\,\lambda\ne0$) the functions of the form $(V(x)-
\lambda)^\varkappa$ will not be used.

The notation is adopted,
$$\alpha(x;\lambda)={5\over{16}}{{(V'(x))^2}\over{(V(x)-\lambda)^{5/2}}}-
{1\over4}{{V''(x)}\over{(V(x)-\lambda)^{3/2}}},\qquad\alpha(x)=\alpha(x;0),$$
$$\hat\alpha(x;\lambda)={5\over{16}}{{|V'(x)|^2}\over{|V(x)-\lambda|^{5/2}}}+
{1\over4}{{|V''(x)|}\over{|V(x)-\lambda|^{3/2}}},\qquad\hat\alpha(x)= \hat\alpha(x;0),$$
$$\xi_{\uparrow\downarrow}(x;\lambda)=\pm\int\limits_{\pm R_1}^x\sqrt{V(x_1)-\lambda}\,dx_1,
\qquad R_1=\begin{cases} R_0,&\lambda\leqslant0,\\ R_2,& {\rm{Im}}\,\lambda\ne0,
\end{cases}$$
$$\xi_{\uparrow\downarrow}(x)=\pm\int\limits_{\pm R_0}^x\sqrt{|V(x_1)|}\,dx_1,\quad
I_{1,\uparrow\downarrow}(x;\lambda)=\pm\int\limits_x^{\pm\infty}\hat\alpha(x_1;\lambda)\, dx_1,\quad
I_{1,\uparrow\downarrow}(x)=I_{1,\uparrow\downarrow}(x;0),$$
$$I_{2,\uparrow\downarrow}(x;\lambda)=\pm\int\limits_x^{\pm\infty}\hat\alpha(x_1;\lambda)
e^{-2{\rm{Re}}(\xi_{\uparrow\downarrow}(x_1;\lambda)-\xi_{\uparrow\downarrow}(x;\lambda))}\,dx_1,$$
$$I_{3,\uparrow\downarrow}(x;\lambda)=\pm\int\limits_{\pm R_1}^x\hat\alpha(x_1;\lambda)
e^{-2{\rm{Re}}(\xi_{\uparrow\downarrow}(x;\lambda)-\xi_{\uparrow\downarrow}(x_1;\lambda))}\,dx_1,$$
$$C=\max\limits_{|x|\geqslant R_0}{{|{\rm{Im}}\,V(x)|}\over{|{\rm{Re}}\,V(x)|}}.$$

The notion of a formal associated function, used in this paper, is defined as follows.

{\bf Definition 3 (1).} The function $\psi_{n,i}(x)$ is called a formal associated function of $i$-th order of
the Hamiltonian $h$ for a spectral value $\lambda_n$, if \be (h-\lambda_n)^{i+1}\psi_{n,i}\equiv0,\qquad
(h-\lambda_n)^{i}\psi_{n,i}\not\equiv 0, \label{canbas1}\ee where the adjective 'formal' emphasizes that a
related function is not necessarily normalizable.
\bigskip

In particular, the associated function of zero order $\psi_{n,0}$ is a formal eigenfunction of $h$ (a solution
of the homogeneous Schr\"odinger equation, not necessarily normalizable).

In this paper we employ the normalizability of functions and in particular the normalizability at $+\infty$ (at
$-\infty$), which is defined as follows.

{\bf Definition 4 (3).} A function $f(x)$ is called normalizable at $+\infty$ (at $-\infty$), if there is $R_+$
($R_-$) such that \be \int\limits_{R_+}^{+\infty}|f(x)|^2\,dx<+\infty\qquad
\bigg(\int\limits_{-\infty}^{R_-}|f(x)|^2\,dx<+\infty\bigg).\ee Otherwise $f(x)$ is called non-normalizable at
$+\infty$ (at $-\infty$).

\section{Estimates on potentials and asymptotics of useful integrals}

{\bf Lemma 5.} {\it If $V(x)\in\cal K$, $\lambda\in\mathbb C$ and either $\lambda\leqslant 0$ or
${\rm{Im}}\,\lambda\ne0$, then there are constants $C_1$, $C_2$, $C_3>0$ such that for any $|x|\geqslant R_1$
the inequalities are valid,
\begin{equation} C_1\leqslant {{|V(x)|^2}\over{|V(x)-\lambda|^2}}\leqslant C_2,\la{a4}\end{equation}
\begin{equation}{\rm{Re}}\,\sqrt{V(x)-\lambda}\geqslant C_3\sqrt{|V(x)|}.\la{a5}\end{equation}
}

{\bf Proof.}

Let us first consider the case $\lambda\leqslant0$. Then the right side of \gl{a4} is obvious. The left side of
\gl{a4} follows from the chain
$${{|V-\lambda|^2}\over{|V|^2}}\equiv{{(1-\lambda/{\rm{Re}}\,V)^2+
{\rm{Im^2}}\,V/{\rm{Re^2}}V}\over{1+{\rm{Im^2}}\,V/{\rm{Re^2}}V}}\leqslant
(1-{\lambda\over\varepsilon})^2+C^2.$$ The inequality \gl{a5} is derived from the chain
$${{|V|}\over{({\rm{Re}}\,\sqrt{V-\lambda})^2}}\equiv
{{2\sqrt{1+ {\rm{Im^2}}\,V/{\rm{Re^2}}V}}\over{1-\lambda/{\rm{Re}}\,V+
\sqrt{(1-\lambda/{\rm{Re}}\,V)^2+{\rm{Im^2}}\,V/{\rm{Re^2}}\,V}}}\leqslant{{2 \sqrt{1+C^2}}\over{1+1}}.$$

Let us now consider the case ${\rm{Im}}\,\lambda\ne0$. In this case the left side of \gl{a4} is provided by the
chain
$${{|V-\lambda|^2}\over{|V|^2}}\equiv{{(1-{\rm{Re}}\,\lambda/
{\rm{Re}}\,V)^2+({\rm{Im}}\,V/{\rm{Re}}V-{\rm{Im}}\,\lambda/
{\rm{Re}}V)^2}\over{1+{\rm{Im^2}}\,V/{\rm{Re^2}}V}}\leqslant
(1+{{|{\rm{Re}}\,\lambda|}\over\varepsilon})^2+(C+{{|{\rm{Im}}\,\lambda|} \over\varepsilon})^2.$$ The right side
of \gl{a4} in the subcase ${\rm{Re}}\,V\geqslant2|{\rm{Re}}\,\lambda|$ follows from the sequence of inequalities
$${{|V|^2}\over{|V-\lambda|^2}}\equiv{{1+{\rm{Im^2}}\,V/{\rm{Re^2}}V}
\over{(1-{\rm{Re}}\,\lambda/{\rm{Re}}\,V)^2+({\rm{Im}}\,V/ {\rm{Re}}V-{\rm{Im}}\,\lambda/{\rm{Re}}V)^2}}$$
$$\leqslant{{1+C^2}\over {(1-|{\rm{Re}}\,\lambda|/{\rm{Re}}\,V)^2}}\leqslant{4(1+C^2)},$$ and in the subcase
${\rm{Re}}\,V\leqslant2|{\rm{Re}}\,\lambda|$ from the sequence\footnote{For the derivation of this chain the
inequality \gl{star} is used.}
$${{|V|^2}\over{|V-\lambda|^2}}\equiv{{1+{\rm{Im^2}}\,V/{\rm{Re^2}}V}
\over{(1-{\rm{Re}}\,\lambda/{\rm{Re}}\,V)^2+({\rm{Im}}\,V/
{\rm{Re}}V-{\rm{Im}}\,\lambda/{\rm{Re}}V)^2}}\leqslant$$ $${{1+C^2}\over
{(1-|{\rm{Re}}\,\lambda|/{\rm{Re}}\,V)^2+(|{\rm{Im}}\,\lambda|/
{\rm{Re}}V-|{\rm{Im}}\,\lambda|/(2|{\rm{Re}}\,\lambda|))^2}}\equiv$$
$${{1+C^2}\over{{{|\lambda|^2}\over{{\rm{Re^2}}\,\lambda}}\Big(
{{|{\rm{Re}}\,\lambda|}\over{{\rm{Re}}\,V}}-{1\over2}-{{{\rm{Re^2}}\,
\lambda}\over{2|\lambda|^2}}\Big)^2+{{{{\rm{Im^2}}\,
\lambda}\over{4|\lambda|^2}}}}}\leqslant{{4|\lambda|^2(1+C^2)}\over {{\rm{Im^2}}\,\lambda}}.$$ One can obtain
\gl{a5} with the help of the inequality
$${{|V|}\over{({\rm{Re}}\,\sqrt{V-\lambda})^2}}\equiv{{2\sqrt{1+
{\rm{Im^2}}\,V/{\rm{Re^2}}V}}\over{1-{\rm{Re}}\,\lambda/{\rm{Re}}\,V+
\sqrt{(1-{\rm{Re}}\,\lambda/{\rm{Re}}\,V)^2+({\rm{Im}}\,V/{\rm{Re}}\,V-
{\rm{Im}}\,\lambda/{\rm{Re}}\,V)^2}}}\leqslant$$
\begin{equation}{{2\sqrt{1+C^2}}\over{1-|{\rm{Re}}\,\lambda|/{\rm{Re}}\,V+
\sqrt{(1-|{\rm{Re}}\,\lambda|/{\rm{Re}}\,V)^2+(|{\rm{Im}}\,V|/{\rm{Re}}\,V-
|{\rm{Im}}\,\lambda|/{\rm{Re}}\,V)^2}}}.\la{a6}\end{equation} In the subcase
${\rm{Re}}\,V\geqslant2|{\rm{Re}}\,\lambda|$ the right side of \gl{a6} is less than or equal to $2\sqrt{1+C^2}$,
wherefrom \gl{a5} follows, and in the subcase ${\rm{Re}}\,V\leqslant2|{\rm{Re}}\,\lambda|$ the right side
\gl{a6} is less than or equal to
$${{2\sqrt{1+C^2}}\over{1-|{\rm{Re}}\,\lambda|/{\rm{Re}}\,V+
\sqrt{(1-|{\rm{Re}}\,\lambda|/{\rm{Re}}\,V)^2+(|{\rm{Im}}\,\lambda|/
{\rm{Re}}\,V-|{\rm{Im}}\,\lambda|/(2{\rm{Re}}\,\lambda))^2}}}\equiv$$
\begin{equation}{{2\sqrt{1+C^2}}\over{1\!-\!|{\rm{Re}}\,\lambda|/{\rm{Re}}\,V\!+\!
(|\lambda|/|{\rm{Re}}\,\lambda|)\sqrt{(|{\rm{Re}}\,\lambda|/{\rm{Re}}\,V\!-\!
1/2\!-\!{\rm{Re^2}}\lambda/(2|\lambda|^2))^2\!+\!{\rm{Re^2}}\lambda
{\rm{Im^2}}\lambda/(4|\lambda|^4)}}}.\la{a7}\end{equation} Insofar as the function
$$f(y)=1-y+
{{|\lambda|}\over{|{\rm{Re}}\,\lambda|}}\sqrt{(y-
{1\over2}-{{{\rm{Re^2}}\lambda}\over{2|\lambda|^2}})^2+{{{\rm{Re^2}}\lambda
{\rm{Im^2}}\lambda}\over{(4|\lambda|^4)}}}$$ has a  minimum at the point $y=1/2+{\rm{Re^2}}\lambda/|\lambda|^2$,
 \gl{a7} is less  than or equal to
$${{2\sqrt{1+C^2}}\over{1-{\rm{Re^2}}\,\lambda/|\lambda|^2+
(|\lambda|/|{\rm{Re}}\,\lambda|)\sqrt{{\rm{Re^4}}\lambda/(4|\lambda|^4)+
{\rm{Re^2}}\lambda{\rm{Im^2}}\lambda/(4|\lambda|^4)}}}\equiv
{{2|\lambda|^2}\over{{\rm{Im^2}}\lambda}}\sqrt{1+C^2}.$$ Thus, Lemma 5 is proved.

{\bf Corollary 1.} If $V(x)\in\cal K$, $\lambda\in\mathbb C$ and either $\lambda\leqslant 0$ or
${\rm{Im}}\,\lambda\ne0$, then
$${{|V'(x)|}\over{|V(x)-\lambda|^{3/2}}}=O\Big({1\over{\xi_{\uparrow\downarrow}(x)}}\Big),
\qquad x\to\pm\infty.$$

{\bf Lemma 6.} {\it If $V(x)\in\cal K$, $\lambda\in\mathbb C$ and either $\lambda\leqslant 0$ or
${\rm{Im}}\,\lambda\ne0$, then for any $|x|\geqslant R_1$ the integrals $I_{1,\uparrow\downarrow}(x;\lambda)$
and $I_{2,\uparrow\downarrow}(x;\lambda)$ converge and the estimates hold:
$$I_{1,\uparrow\downarrow}(x;\lambda)=O\Big({1\over{\xi_{\uparrow\downarrow}(x)}}\Big),\qquad
x\to\pm\infty,$$
$$I_{2,\uparrow\downarrow}(x;\lambda)=O\Big({1\over{\xi_{\uparrow\downarrow}^2(x)}}\Big),\qquad
x\to\pm\infty,$$
$$I_{3,\uparrow\downarrow}(x;\lambda)=O\Big({1\over{\xi_{\uparrow\downarrow}^2(x)}}\Big),\qquad
x\to\pm\infty.$$}

{\bf Proof.}

Because the  proofs for the cases $x\to+\infty$ and $x\to-\infty$ are similar, we shall consider the case
$x\to+\infty$ only. Due to $V\in\cal K$ and Lemma~5 there are positive constants $C_4$, \dots, $C_9$ and $\xi_0$
such that
$$I_{1,\uparrow}(x;\lambda)\leqslant
C_4\!\int\limits_x^{+\infty}\!{{|V'|^2}\over{|V|^{5/2}}}dx_1+
C_5\!\int\limits_x^{+\infty}\!{{|V''|}\over{|V|^{3/2}}}dx_1\leqslant
C_6\!\int\limits_x^{+\infty}\!{\sqrt{|V|}\over{\xi_\uparrow^2}}dx_1\equiv
C_6\!\int\limits_x^{+\infty}\!{{\xi'_\uparrow}\over{\xi_\uparrow^2}}dx_1= {C_6\over{\xi_\uparrow(x)}},$$
$$I_{2,\uparrow}(x;\lambda)\leqslant
C_7\int\limits_x^{+\infty}{{\xi'_\uparrow(x_1)}\over{\xi_\uparrow^2(x_1)}}
e^{-2\int\limits_x^{x_1}{\rm{Re}}\,\sqrt{V(x_2)-\lambda}dx_2} dx_1\leqslant
C_7\int\limits_x^{+\infty}{{\xi'_\uparrow(x_1)}\over{\xi_\uparrow^2(x_1)}}
e^{-C_8\int\limits_x^{x_1}\sqrt{|V(x_2)|}dx_2} dx_1\equiv$$
$$C_7\int\limits_x^{+\infty}{{\xi'_\uparrow(x_1)}\over{\xi_\uparrow^2(x_1)}}
e^{C_8(\xi_\uparrow(x)-\xi_\uparrow(x_1))} dx_1\leqslant
{C_7\over{\xi_\uparrow^2(x)}}e^{C_8\xi_\uparrow(x)}\int\limits_x^{+\infty}
\xi'_\uparrow(x_1)e^{-C_8\xi_\uparrow(x_1)} dx_1={{C_7/C_8}\over{\xi_\uparrow^2(x)}},$$
$$I_{3,\uparrow}(x;\lambda)\leqslant
C_9\int\limits_{R_0}^x{{\xi'_\uparrow(x_1)}\over{(\xi_0+\xi_\uparrow(x_1))^2}}
e^{-C_8(\xi_\uparrow(x)-\xi_\uparrow(x_1))} dx_1=
C_9e^{-C_8\xi_\uparrow(x)}\int\limits_0^{\xi_\uparrow(x)}{{e^{C_8\xi} d\xi}\over{(\xi_0+\xi)^2}} =$$
$$C_9e^{-C_8\xi_\uparrow(x)}\Big(\int\limits_0^{\xi_\uparrow(x)/2}+\int\limits_{
\xi_\uparrow(x)/2}^{\xi_\uparrow(x)}\Big){{e^{C_8\xi} d\xi}\over{(\xi_0+\xi)^2}}=
C_9e^{-C_8\xi_\uparrow(x)}\Big[e^{C_8\xi_\uparrow(x)/2}\Big({1\over\xi_0}-
{1\over{\xi_0+\xi_\uparrow(x)/2}}\Big)+$$
$${1\over{(\xi_0+\xi_\uparrow(x)/2)^2}}{1\over
C_8}\Big(e^{C_8\xi_\uparrow(x)}-e^{C_8\xi_\uparrow(x)/2}\Big)\Big]\leqslant
C_9\Big[{1\over\xi_0}e^{-C_8\xi_\uparrow(x)/2}+{{4/C_8}\over{\xi_\uparrow^2(x)}}\Big],$$ wherefrom Lemma 6
follows.

{\bf Lemma 7.} {\it Let: 1) $V(x)\in\cal K$; 2) $\lambda\in\mathbb C$ and either $\lambda\leqslant 0$ or
${\rm{Im}}\,\lambda\ne0$; 3) the integral
\begin{equation}\int\limits_{R_1}^{+\infty}{{dx_1}\over\sqrt{V(x_1)-\lambda}}\qquad
\Big(\int\limits_{-\infty}^{-R_1}{{dx_1}\over\sqrt{V(x_1)-\lambda}}\Big) \label{a9}\end{equation} converges.
Then the integral
\begin{equation}\int\limits_{R_0}^{+\infty}{{dx_1}\over\sqrt{|V(x_1)|}}\qquad
\Big(\int\limits_{-\infty}^{-R_0}{{dx_1}\over\sqrt{|V(x_1)|}}\Big) \label{a10}\end{equation} converges too,
\be\lim\limits_{x\to\pm\infty}V(x)=\infty\label{a12}\ee and
\begin{equation}{1\over{V(x)}}=O\Big({1\over{\xi_{\uparrow\downarrow}(x)}}\int\limits_x^{\pm
\infty}{{dx_1}\over\sqrt{|V(x_1)|}}\Big),\qquad x\to\pm\infty.\label{a11}\end{equation}}

{\bf Proof.}

As the integral \gl{a9} converges, then the integral
$$\int\limits_{R_1}^{+\infty}{\rm{Re}}{1\over\sqrt{V(x_1)-\lambda}}{dx_1}
\equiv \int\limits_{R_1}^{+\infty}{{{\rm{Re}}\sqrt{V(x_1)-\lambda}}
\over{|V(x_1)-\lambda|}}{dx_1}\qquad\bigg(\int\limits_{-\infty}^{-R_1} {{{\rm{Re}}\sqrt{V(x_1)-\lambda}}
\over{|V(x_1)-\lambda|}}{dx_1}\bigg)$$ converges too. In view of Lemma 5 there are constants $C_1$ and $C_3$
such that ${\rm{Re}}\sqrt{V(x)-\lambda}\geqslant C_3\sqrt{|V(x)|}$ and
$|V(x)-\lambda|\leqslant{1\over\sqrt{C_1}}|V(x)|$ for any $|x|\geqslant R_1$. Hence the integral
$$\int\limits_{R_1}^{+\infty}{{dx_1}\over\sqrt{|V(x_1)|}}\leqslant
{1\over{\sqrt{C_1}C_3}} \int\limits_{R_1}^{+\infty}{{{\rm{Re}}\sqrt{V(x_1)-\lambda}}
\over{|V(x_1)-\lambda|}}{dx_1}$$
$$\bigg(\int\limits_{-\infty}^{-R_1}{{dx_1}\over\sqrt{|V(x_1)|}}\leqslant
{1\over{\sqrt{C_1}C_3}} \int\limits_{-\infty}^{-R_1}{{{\rm{Re}}\sqrt{V(x_1)-\lambda}}
\over{|V(x_1)-\lambda|}}{dx_1}\bigg)$$ converges  as much as the integral \gl{a10}.

Let us now  check \gl{a12} in the case $x\to+\infty$ (examination of the case $x\to-\infty$ is similar). The
integral
$$\int\limits_{R_1}^{+\infty}{{V'(x_1)}\over{(V(x_1)-\lambda)^2}}dx_1$$
 converges owing to convergence of \gl{a10} and boundedness of
$|V'|/|V-\lambda|^{3/2}$ for $x\geqslant R_1$ (see corollary~1). Hence the limit of the function
$${1\over{V(x)-\lambda}}={1\over{V(R_1)-\lambda}}-
\int\limits_{R_1}^x{{V'(x_1)}\over{(V(x_1)-\lambda)^2}}dx_1, \qquad x\geqslant R_1$$ for $x\to+\infty$ is
finite. Moreover, because of convergence of \gl{a9} this limit  is zero. Thus \gl{a12} holds.

Validity of \gl{a11} for $x\to+\infty$ (consideration of the case $x\to-\infty$ is similar) is justified by the
fact that for $V\in\cal K$ there is $C_4>0$ such that $|V'|/|V|^{3/2}\leqslant C_4/\xi_\uparrow$ for any
$x\geqslant R_0$ and by the chain
$${1\over{|V(x)|}}=\Big|\int\limits_x^{+\infty}{{V'(x_1)}\over
{V^2(x_1)}}dx_1\Big|\leqslant C_4\int\limits_x^{+\infty}{{dx_1}\over
{\sqrt{|V(x_1)|}\xi_\uparrow(x_1)}}\leqslant{C_4\over{\xi_\uparrow(x)}}\int\limits_x^{+\infty} {{dx_1}\over
{\sqrt{|V(x_1)|}}}.$$

\section{Asymptotics of formal eigenfunctions of a Hamiltonian}

Asymptotic behavior of formal eigenfunctions of a Hamiltonian with potential
belonging to $\cal K$ is described by the\\

{\bf Lemma 8.} {\it Let: 1) $V(x)\in\cal K$; 2) $\lambda\in\mathbb C$ and either $\lambda\leqslant 0$ or
${\rm{Im}}\,\lambda\ne0$. Then there are functions $\varphi_{0,\uparrow\downarrow}(x)$ normalizable at
$\pm\infty$ being zero-modes of $h-\lambda$ and functions $\hat\varphi_{0,\uparrow\downarrow}(x)$
non-normalizable at $\pm\infty$ being zero-modes of $h-\lambda$ such that\footnote{Leading terms of asymptotics
\gl{a123} and \gl{a14} are well known (see for example \cite{shubin}).}
\begin{equation}\varphi_{0,\uparrow\downarrow}(x)={1\over\root4\of{V(x)-\lambda}}e^{
-\xi_{\uparrow\downarrow}(x;\lambda)}\Big[1-{1\over2}\int\limits_x^{\pm\infty}
\alpha(x_1;\lambda)\,dx_1+O\Big({1\over{\xi_{\uparrow\downarrow}^2(x)}}\Big)\Big],\qquad
x\to\pm\infty,\la{a123}\end{equation}
\begin{equation}{{\varphi'_{0,\uparrow\downarrow}(x)}\over{\varphi_{0,\uparrow\downarrow}(x)}}=
\mp\sqrt{V(x)-\lambda} \Big[1\pm{1\over4}{{V'(x)}\over{(V(x)-\lambda)^{3/2}}}+
O\Big({1\over{\xi_{\uparrow\downarrow}^2(x)}}\Big)\Big],\qquad x\to\pm\infty,\la{a13}\end{equation}
\begin{equation}\hat\varphi_{0,\uparrow\downarrow}(x)={1\over\root4\of{V(x)-\lambda}}e^{
\xi_{\uparrow\downarrow}(x;\lambda)}\Big[1+{1\over2}\int\limits_x^{\pm\infty}
\alpha(x_1;\lambda)\,dx_1+O\Big({1\over{\xi_{\uparrow\downarrow}^2(x)}}\Big)\Big],\qquad
x\to\pm\infty,\la{a14}\end{equation}
\begin{equation}{{\hat\varphi'_{0,\uparrow\downarrow}(x)}\over
{\hat\varphi_{0,\uparrow\downarrow}(x)}}=\pm\sqrt{V(x)-\lambda}
\Big[1\mp{1\over4}{{V'(x)}\over{(V(x)-\lambda)^{3/2}}}+
O\Big({1\over{\xi_{\uparrow\downarrow}^2(x)}}\Big)\Big],\qquad x\to\pm\infty.\la{a15}\end{equation}}

{\bf Proof.}

We shall consider the case $x\to+\infty$ only because examination of the case $x\to-\infty$ is analogous. Let us
show that the series
$$\varphi_{0,\uparrow}(x)={1\over\root4\of{V(x)-\lambda}}
\sum\limits_{n=0}^{+\infty}\int\limits_x^{+\infty}dx_1\,{\rm{sh}\,}
(\xi_\uparrow(x;\lambda)-\xi_\uparrow(x_1;\lambda))\alpha(x_1;\lambda)$$
$$\times \int\limits_{x_1}^{
+\infty}dx_2\,{\rm{sh}\,}(\xi_\uparrow(x_1;\lambda)-\xi_\uparrow(x_2;\lambda))
\alpha(x_2;\lambda)$$\begin{equation}\ldots \int\limits_{x_{n-1}}^{
+\infty}dx_n\,{\rm{sh}\,}(\xi_\uparrow(x_{n-1};\lambda)-\xi_\uparrow(x_n;\lambda))
\alpha(x_n;\lambda)e^{-\xi_\uparrow(x_n\lambda)}\la{a16}\end{equation} converges and gives the required function
$\varphi_{0,\uparrow}(x)$. Convergence of \gl{a16} is provided by the fact that the series \gl{a16} is majorized
by the series
$$\sum\limits_{n=0}^{+\infty}\int\limits_x^{+\infty}dx_1\,e^{{\rm{Re}\,}
(\xi_\uparrow(x_1;\lambda)-\xi_\uparrow(x;\lambda))}\hat\alpha(x_1;\lambda)\int\limits_{x_1}^{
+\infty}dx_2\,e^{{\rm{Re}\,}(\xi_\uparrow(x_2;\lambda)-\xi_\uparrow(x_1;\lambda))}
\hat\alpha(x_2;\lambda)\ldots$$
$$\int\limits_{x_{n-1}}^{
+\infty}dx_n\,e^{{\rm{Re}\,}(\xi_\uparrow(x_n;\lambda)-\xi_\uparrow(x_{n-1};\lambda))}
\hat\alpha(x_n;\lambda)e^{-{\rm{Re}\,}\xi_\uparrow(x_n;\lambda)}\equiv$$
$$e^{-{\rm{Re}\,}\xi_\uparrow(x;\lambda)}\sum\limits_{n=0}^{+\infty}
\int\limits_x^{+\infty}dx_1\,\hat\alpha(x_1;\lambda)\int\limits_{x_1}^{
+\infty}dx_2\,\hat\alpha(x_2;\lambda)\ldots\int\limits_{x_{n-1}}^{ +\infty}dx_n\,\hat\alpha(x_n;\lambda)=$$
$$e^{-{\rm{Re}\,}\xi_\uparrow(x;\lambda)}\sum\limits_{n=0}^{+\infty}{1\over{n!}}
\Big(\int\limits_x^{+\infty}\hat\alpha(x_1;\lambda),dx_1\Big)^n\equiv
e^{-{\rm{Re}\,}\xi_\uparrow(x;\lambda)+I_{1,\uparrow}(x;\lambda)}.$$ From this estimate it follows also that due
to Lemma~6 the asymptotics \gl{a123} for the function \gl{a16}  is valid. Insofar as the series of first and
second derivatives of \gl{a16} are majorized for $x$ belonging to any segment $[x_1,x_2]\subset[R_1,+\infty[$ by
the  series independent of $x$,
$$\sum\limits_{n=0}^{+\infty}\max\limits_{[x_1,x_2]}|\xi'_\uparrow(x;\lambda)|
{1\over{n!}}(I_{1,\uparrow}(R_1;\lambda))^n$$ and
$$\sum\limits_{n=0}^{+\infty}\Big\{\max\limits_{[x_1,x_2]}[|\xi''_\uparrow(x;\lambda)|
+|\xi'_\uparrow(x;\lambda)|^2]{1\over{n!}}(I_{1,\uparrow}(R_1;\lambda))^n+$$
$$\max\limits_{[x_1,x_2]}[|\alpha(x;\lambda)\xi'_\uparrow(x;\lambda)|]
{1\over{(n-1)!}}(I_{1,\uparrow}(R_1;\lambda))^{n-1}\Big\},$$ it is possible to differentiate twice the series
\gl{a16} term by term. Calculation of $h-\lambda$ applied to the function \gl{a16} allows to check that this
function is a zero-mode of $h-\lambda$. One can check also \gl{a13}, with the help of \gl{a16}, using Lemma~6,
corollary~1 and the fact that the absolute value of the derivative of $n$-th term in \gl{a16}  is less than or
equal to
$${1\over{n!}}\sqrt{|V(x)-\lambda|}(I_{1,\uparrow}(x;\lambda))^ne^{-{\rm{Re}\,}
\xi_\uparrow(x;\lambda)}.$$ To prove normalizability of $\varphi_{0,\uparrow}(x)$ at $+\infty$ it is sufficient
to prove normalizability at $+\infty$ of the leading term of the asymptotics \gl{a123}. The latter comes out of
the fact that for $V(x)\in\cal K$  the chain of inequalities holds  due to Lemma~5,
$${e^{-2{\rm{Re}\,}\xi_\uparrow(x;\lambda)}\over\sqrt{|V(x)-\lambda|}}\leqslant
{{\root4\of{C_2}}\over\sqrt{|V(x)|}}e^{-2C_3\int\limits_{R_1}^x\sqrt{
|V(x_1)|}\,dx_1}\leqslant{{\root4\of{C_2}}\over\sqrt\varepsilon}e^{-2C_3
\int\limits_{R_1}^x\sqrt{\varepsilon}\,dx_1}\leqslant{{\root4\of{C_2}}\over
\sqrt\varepsilon}e^{-2C_3\sqrt{\varepsilon}(x-R_1)},$$ the right side of which is obviously normalizable at
$+\infty$.

Let us prove now that the required function $\hat\varphi_{0,\uparrow}(x)$ can be written in the form
\begin{equation}\hat\varphi_{0,\uparrow}(x)=2\varphi_{0,\uparrow}(x)\int\limits_{R_3}^x
{{dx_1}\over{\varphi_{0,\uparrow}^2(x_1)}},\la{a17}\end{equation} where $R_3\geqslant R_1$ is a constant such
that $\varphi_{0,\uparrow}(x)$ has no zeroes for $x\geqslant R_3$ (existence of $R_3$ is obvious because
of~\gl{a123}). The fact that the function \gl{a17} is a zero-mode of $h-\lambda$ follows from elementary
calculations. To prove that the asymptotics \gl{a14} and \gl{a15} for the function \gl{a17} are valid it is
sufficient to prove that
$$\int\limits_{R_3}^x{{dx_1}\over{\varphi_{0,\uparrow}^2(x_1)}}={1\over2}
e^{2\xi_\uparrow(x;\lambda)}\Big[1+\int\limits_x^{+\infty}\alpha(x_1;\lambda)\,
dx_1+\Big({1\over{\xi_\uparrow^2(x)}}\Big)\Big]$$ in view of \gl{a123}, \gl{a13}, \gl{a17} and the obvious
formula
$${{\hat\varphi'_{0,\uparrow}(x)}\over{\hat\varphi_{0,\uparrow}(x)}}={{\varphi'_{0,\uparrow}(x)}
\over{\varphi_{0,\uparrow}(x)}}+{1\over{\varphi^2_{0,\uparrow}(x)\int\limits_{R_3}^x
{{dx_1}\over{\varphi_{0,\uparrow}^2(x_1)}}}}.$$ By virtue of \gl{a123}
\begin{equation}{1\over{\varphi^2_{0,\uparrow}(x)}}=\sqrt{V(x)-\lambda}\,\,
e^{2\xi_\uparrow(x;\lambda)}\Big[1+\int\limits_x^{+\infty}\alpha(x_1;\lambda)\,
dx_1+\Big({1\over{\xi_\uparrow^2(x)}}\Big)\Big].\la{a18}\end{equation} Because of Lemma 6 the contribution of
first and second terms of the right side of \gl{a18} at $\int_{R_3}^xdx_1/\varphi^2_{0,\uparrow}(x_1)$  is given
by
$$\int\limits_{R_3}^x\sqrt{V(x_1)-\lambda}\,\,e^{2\xi_\uparrow(x_1;\lambda)}
\Big[1+\int\limits_{x_1}^{+\infty}\alpha(x_2;\lambda)\,dx_2\Big]dx_1=$$
$${1\over2}\int\limits_{R_3}^x\Big(e^{2\xi_\uparrow(x_1;\lambda)}\Big)'
\Big[1+\int\limits_{x_1}^{+\infty}\alpha(x_2;\lambda)\,dx_2\Big]dx_1=$$
$${1\over2}\Big\{e^{2\xi_\uparrow(x;\lambda)}\Big[1+\int\limits_x^{+\infty}
\alpha(x_2;\lambda)\,dx_2\Big]+O(1)+\int\limits_{R_3}^x e^{2\xi_\uparrow(x_1;\lambda)}
\alpha(x_1;\lambda)\,dx_1\Big\}=$$
$$={1\over2}e^{2\xi_\uparrow(x;\lambda)}\Big[1+\int\limits_x^{+\infty}
\alpha(x_1;\lambda)\,dx_1+\Big({1\over{\xi_\uparrow^2(x)}}\Big)\Big],\qquad x\to+\infty.$$ Due to local
boundedness of $1/\varphi_{0,\uparrow}^2(x)$ for $x\geqslant R_3$ the contribution of the third term of right
side of \gl{a18}  is less than or equal to the integral
$$C_4\int\limits_{R_3}^x{\sqrt{|V(x_1)-\lambda|}\over{(\xi_0+\xi_\uparrow(x_1))^2}}
e^{2{\rm{Re}\,}\xi_\uparrow(x_1;\lambda)}dx_1\equiv$$
\begin{equation}C_4e^{2{\rm{Re}\,}\xi_\uparrow(x;\lambda)}\int\limits_{R_3}^x{\sqrt{|V(x_1)-
\lambda|}\over{(\xi_0+\xi_\uparrow(x_1))^2}}e^{-2\int\limits_{x_1}^x{\rm{Re}\,}
\sqrt{V(x_2)-\lambda}\,dx_2}dx_1,\la{a19}\end{equation} where $C_4$ and $\xi_0$ are  positive constants. For
some positive constants $C_5$ and $C_6$ the integral \gl{a19}, in view of Lemma~5, is less than or equal  to the
integral
$$C_5e^{2{\rm{Re}\,}\xi_\uparrow(x;\lambda)}\int\limits_{R_0}^x{\sqrt{|V(x_1)|}
\over{(\xi_0+\xi_\uparrow(x_1))^2}}e^{-C_6(\xi_\uparrow(x)-\xi_\uparrow(x_1))}dx_1\equiv$$
$$C_5e^{2{\rm{Re}\,}\xi_\uparrow(x;\lambda)}\int\limits_{R_0}^x{{\xi'_\uparrow(x_1)}
\over{(\xi_0+\xi_\uparrow(x_1))^2}}e^{-C_6(\xi_\uparrow(x)-\xi_\uparrow(x_1))}dx_1,$$ which is equal (see the
proof of Lemma 6) to $O(e^{2{\rm{Re}\,}\xi_\uparrow(x;\lambda)}/\xi_\uparrow^2(x))$, $x\to+\infty$. Thus,
\gl{a14} and \gl{a15} hold.

Finally in order to prove non-normalizability at $+\infty$ of function \gl{a17} let us first  prove the
auxiliary inequality
\begin{equation}|V(x)|\leqslant
C_0(\xi_0+\xi_\uparrow(x))^\gamma,\qquad x\geqslant R_0,\la{a20}\end{equation} where $C_0$ and $\gamma$ are some
positive constants. This equality for $V(x)\in\cal K$ follows from the sequence
$$|V(x)|=|V(R_0)e^{\ln(V(x)/V(R_0))}|\leqslant|V(R_0)|e^{\int\limits_{R_0}^x
{{|V'(x_1)|}\over{|V(x_1)|}}dx_1}\leqslant$$
$$|V(R_0)|e^{\gamma\int\limits_{R_0}^x
{\sqrt{|V(x_1)|}\over{\xi_0+\xi_\uparrow(x_1)}}dx_1}\equiv|V(R_0)|e^{\gamma \int\limits_{R_0}^x
{{\xi'_\uparrow(x_1)}\over{\xi_0+\xi_\uparrow(x_1)}}dx_1}=
|V(R_0)|{{(\xi_0+\xi_\uparrow(x))^\gamma}\over{(\xi_0+\xi_\uparrow(R_0))^\gamma}}.$$ To prove
non-normalizability at $+\infty$ of function \gl{a17} it is sufficient to prove non-nor\-ma\-li\-za\-bi\-li\-ty
at $+\infty$ of the leading term of asymptotics \gl{a14}. The latter is provided by the fact that in view of
\gl{a20} and Lemma~5 the chain holds,
$${e^{2{\rm{Re}\,}\xi_\uparrow(x;\lambda)}\over\sqrt{|V(x)-\lambda|}}\geqslant
{{\root4\of{C_1}}\over\sqrt{|V(x)|}}e^{2C_3\int\limits_{R_1}^x\sqrt{ |V(x_1)|}\,dx_1}\equiv{{\root4\of{C_1}
\xi'_\uparrow(x_1)}\over{|V(x)|}}e^{2C_3(\xi_\uparrow(x)-\xi_\uparrow(R_0))}\geqslant$$
$${{\root4\of{C_2}}\over{C_0}}{{\xi'_\uparrow(x)}\over{(\xi_0+\xi_\uparrow(x))^\gamma}}
e^{2C_3(\xi_\uparrow(x)-\xi_\uparrow(R_0))},$$ the right side of which is  non-normalizable at $+\infty$.
Lemma~8 is proved.

{\bf Remark 1.} The asymptotics \gl{a14} and \gl{a15} are valid for any  zero-mode of $h-\lambda$ linear
independent of $\varphi_{0,\uparrow\downarrow}(x)$ (after its proper
 normalization).

{\bf Corollary 2.} Let Hamiltonians $h^\pm=-\partial^2+V_{1,2}(x)$ be intertwined by
$q_1^\pm=\mp\partial+\chi(x)$:
$$q_1^\pm h^\mp=h^\pm q_1^\pm,\qquad h^\pm=q_1^\pm
q_1^\mp+\lambda=\lambda+\chi^2\mp\chi'.$$ Suppose also $\varphi(x)$ to be a zero-mode of $q_1^-$ so that
$\chi=-\varphi'/\varphi$. Then
$$\triangle V\equiv V_2-V_1=2\chi'\equiv-2(\ln\varphi)''\equiv
-2\Big[{\varphi''\over\varphi}-\Big({\varphi'\over\varphi}\Big)^2\Big]\equiv
-2\Big[V_1-\lambda-\Big({\varphi'\over\varphi}\Big)^2\Big].$$ At last suppose that $V_1(x)\in\cal K$,
$\lambda\in\mathbb C$ and either $\lambda\leqslant0$ or ${\rm{Im}\,}\lambda\ne0$. Then because of~\gl{a13},
\gl{a15}, Lemma~5 and Corollary~1
\begin{equation}\triangle
V=\pm{{V'_1(x)}\over\sqrt{V_1(x)-\lambda}}+O\Big({{V_1(x)}\over{\xi_\uparrow^2(x)}} \Big), \qquad
x\to+\infty,\la{a21}\end{equation} if  at $x\to+\infty$ the asymptotics \gl{a123} or respectively \gl{a14} is
valid for $\varphi$. For the case when at $x\to-\infty$ the asymptotics \gl{a123} or respectively \gl{a14} is
valid for $\varphi$,
\begin{equation}\triangle
V=\mp{{V'_1(x)}\over\sqrt{V_1(x)-\lambda}}+O\Big({{V_1(x)}\over{\xi_\downarrow^2(x)}} \Big), \qquad
x\to-\infty.\la{a22}\end{equation} Finally,
\begin{equation}\triangle V'\equiv-2\Big\{V'_1-2{\varphi'\over\varphi}\Big[{\varphi''
\over\varphi}-\Big({\varphi'\over\varphi}\Big)^2\Big]\Big\}\equiv-2\Big[ V'_1+{\varphi'\over\varphi}\triangle
V\Big]=O\Big({{V_1^{3/2}(x)}\over{\xi_{\uparrow\downarrow}^2(x)}}\Big),\qquad
x\to\pm\infty,\la{a23}\end{equation}
$$\triangle V''\equiv-2\Big\{V''_1+\triangle V\Big[{\varphi''
\over\varphi}-\Big({\varphi'\over\varphi}\Big)^2\Big]+{\varphi'\over\varphi} \triangle V'\Big\}\equiv$$
\begin{equation}-2\Big[V''_1-{1\over2}\triangle V^2+{\varphi'\over \varphi}\triangle
V'\Big]=O\Big({{V_1^2(x)}\over{\xi_{\uparrow\downarrow}^2(x)}}\Big),\qquad x\to\pm\infty,\la{a24}\end{equation}
independently of asymptotics of $\varphi$.

It also follows from \gl{a21}, \gl{a22}, Lemma~5 and Corollary~1 that
$$\triangle
V=O\Big({{V_1(x)-\lambda}\over{\xi_\uparrow(x)}}\Big)=o(V_1-\lambda)=o(V_1),\qquad x\to\pm\infty,$$ {\it i.e.}
that
\begin{equation}V_2-\lambda=V_1-\lambda+o(V_1-\lambda)=(V_1-\lambda)[1+o(1)],
\qquad x\to\pm\infty,\la{a25}\end{equation}
\begin{equation}V_2=V_1+o(V_1)=V_1[1+o(1)],
\qquad x\to\pm\infty.\la{a26}\end{equation}

{\bf Corollary 3 (1).} There are no  degenerate eigenvalues of the Hamiltonian  with a potential belonging to
$K$, satisfying either $\lambda\leqslant0$ or ${\rm{Im}}\,\lambda\ne0$, {\it i.e.} eigenvalues, whose geometric
multiplicity is more than 1 (eigenvalues, for which there are more than one linearly independent eigenfunction).
Hence, for the Hamiltonian with a potential belonging to $K$ there are no more than one Jordan cell made of an
eigenfunction and associated functions, normalizable on the whole axis, for any given eigenvalue $\lambda$ such
that either $\lambda\leqslant0$ or ${\rm{Im}}\,\lambda\ne0$.
\section{Asymptotics of formal associated functions of a Hamiltonian}

The asymptotic behavior of formal associated functions of a Hamiltonian
$h$ with a potential belonging to $\cal K$ is characterized by \\

{\bf Lemma 9.} {\it Let: 1) $h=-\partial^2+V(x)$, $V(x)\in\cal K$; 2) $\lambda\in\mathbb C$ and either
$\lambda\leqslant0$ or ${\rm Im}\,\lambda\ne0$; 3) $\eta_{\uparrow\downarrow}(x)=\pm\int\limits_{\pm
R_0}^xdx_1/\sqrt{|V(x_1)|}$. Then there are denumerable sequences:\qquad $\varphi_{n,\uparrow\downarrow}(x)$ of
formal associated functions of $h$ for a spectral value $\lambda$, normalizable at $\pm\infty$, and \qquad
$\hat\varphi_{n,\uparrow\downarrow}(x)$ of formal
associated functions, non-normalizable at $\pm\infty$,\\
 such that:
\be h\varphi_{0,\uparrow\downarrow}=\lambda\varphi_{0,\uparrow\downarrow},\qquad(h-\lambda)
\varphi_{n,\uparrow\downarrow}=\varphi_{n-1,\uparrow\downarrow},\quad n\geqslant1,\la{2.1}\ee \be
h\hat\varphi_{0,\uparrow\downarrow}=\lambda\hat\varphi_{0,\uparrow\downarrow},\qquad(h-\lambda)
\hat\varphi_{n,\uparrow\downarrow}=\hat\varphi_{n-1,\uparrow\downarrow},\quad n\geqslant1;\la{2.2} \ee if
$\pm\int\limits_{\pm R_0}^{\pm\infty}dx_1/\sqrt{|V(x_1)|}<+\infty$, then for $ x\to\pm\infty,$
\be\varphi_{n,\uparrow\downarrow}(x)={1\over{n!\root {4\,}\of{V(x)-\lambda}}}
\bigg(\pm{1\over2}\int\limits_{\pm\infty}^x{{dx_1}\over\sqrt{V(x_1)-
\lambda}}\bigg)^ne^{-\xi_{\uparrow\downarrow}(x;\lambda)}\bigg[1+O\bigg({1\over{\xi_{\uparrow\downarrow}(x)}}
\bigg)\bigg],\la{2.3}\ee \be\hat\varphi_{n,\uparrow\downarrow}(x)={1\over{n!\root {4\,}\of{V(x)-\lambda}}}
\bigg(\mp{1\over2}\int\limits_{\pm\infty}^x{{dx_1}\over\sqrt{V(x_1)-
\lambda}}\bigg)^ne^{\xi_{\uparrow\downarrow}(x;\lambda)}\bigg[1+O\bigg({1\over{\xi_{\uparrow\downarrow}(x)}}
\bigg)\bigg],\la{2.4}\ee \be\varphi'_{n,\uparrow\downarrow}(x)=\mp{1\over{n!}}\root4\of{V(x)-\lambda}\bigg(\pm
{1\over2}\int\limits_{\pm\infty}^x{{dx_1}\over\sqrt{V(x_1)-\lambda}}\bigg)^n
e^{-\xi_{\uparrow\downarrow}(x;\lambda)}\bigg[1+O\bigg({1\over{\xi_{\uparrow\downarrow}(x)}}\bigg)\bigg]
\la{2.5}\ee and if $\pm\int\limits_{\pm R_0}^{\pm\infty}dx_1/\sqrt{|V(x_1)|}=+\infty$, then
\be\varphi_{n,\uparrow\downarrow}(x)={1\over{n!\root {4\,}\of{V(x)-\lambda}}} \bigg(\pm{1\over2}\int\limits_{\pm
R_1}^x{{dx_1}\over\sqrt{V(x_1)-
\lambda}}\bigg)^ne^{-\xi_{\uparrow\downarrow}(x;\lambda)}\bigg[1+O\bigg({{\ln\eta_{\uparrow\downarrow}(x)
}\over{\eta_{\uparrow\downarrow}(x)}} \bigg)\bigg],\la{2.6}\ee
\be\hat\varphi_{n,\uparrow\downarrow}(x)={1\over{n!\root {4\,}\of{V(x)-\lambda}}}
\bigg(\mp{1\over2}\int\limits_{\pm R_1}^x{{dx_1}\over\sqrt{V(x_1)-
\lambda}}\bigg)^ne^{\xi_{\uparrow\downarrow}(x;\lambda)}\bigg[1+O\bigg({{\ln\eta_{\uparrow\downarrow}(x)
}\over{\eta_{\uparrow\downarrow}(x)}} \bigg)\bigg],\la{2.7}\ee
\be\varphi'_{n,\uparrow\downarrow}(x)\!=\!\mp{1\over{n!}}\root4\of{V(x)\!-\!\lambda}\bigg(\pm
{1\over2}\!\!\!\int\limits_{\pm R_1}^x\!\!\!{{dx_1}\over\sqrt{V(x_1)-\lambda}}\bigg)^n
e^{-\xi_{\uparrow\downarrow}(x;\lambda)}\bigg[1+O\bigg({{\ln\eta_{\uparrow\downarrow}(x)}
\over{\eta_{\uparrow\downarrow}(x)}}\bigg)\bigg].\la{2.8}\ee}

{\bf Proof.}

Let us  prove the existence of $\varphi_{n,\uparrow}(x)$ and $\hat\varphi_{n,\uparrow}(x)$ only, because the
proof of existence of $\varphi_{n,\downarrow}(x)$ and $\hat\varphi_{n,\downarrow}(x)$ is analogous. The
existence of $\varphi_{0,\uparrow}(x)$ and $\hat\varphi_{0,\uparrow}(x)$ was proved in Lemma~8 and  in view of
$V(x)\in\cal K$ the  estimate $O(1/\xi_\uparrow(x))= O(\ln\eta_\uparrow(x)/\eta_\uparrow(x))$, $x\to+\infty$
(cf. Lemma~9 and Lemma~8)  follows  from the chain
\begin{equation}\eta_\uparrow(x)=\int\limits_{R_0}^x{{dx_1}\over\sqrt{|V(x_1)|}}\leqslant
{1\over\varepsilon}\int\limits_{R_0}^x\sqrt{|V(x_1)|}\,dx_1={1\over\varepsilon}
\xi_\uparrow(x).\la{a29'}\end{equation}

Suppose now the existence of $\varphi_{l,\uparrow}(x)$ and $\hat\varphi_{l,\uparrow}(x)$ and let us prove the
existence of $\varphi_{l+1,\uparrow}(x)$ and $\hat\varphi_{l+1,\uparrow}(x)$. In this way the Lemma will be
completely proved.

Consider  the case $\int\limits_{R_0}^{+\infty}dx_1/ \sqrt{|V(x_1)|}<+\infty$. One can check that in this case
$\varphi_{l+1,\uparrow}(x)$ and $\hat\varphi_{l+1,\uparrow}(x)$ can be written in the form
\begin{equation}\varphi_{l+1,\uparrow}(x)=-{1\over2}\Big\{\hat\varphi_{0,\uparrow}(x)\int
\limits_{+\infty}^x\varphi_{0,\uparrow}(x_1)\varphi_{l,\uparrow}(x_1)\,dx_1-\varphi_{0,\uparrow}(x)
\int\limits_{+\infty}^x\hat\varphi_{0,\uparrow}(x_1)\varphi_{l,\uparrow}(x_1)\,dx_1\Big\},
\la{a30}\end{equation}
\begin{equation}\hat\varphi_{l+1,\uparrow}(x)=-{1\over2}\Big\{\hat\varphi_{0,\uparrow}(x)\int
\limits_{+\infty}^x\varphi_{0,\uparrow}(x_1)\hat\varphi_{l,\uparrow}(x_1)\,dx_1-\varphi_{0,\uparrow}(x)
\int\limits_{R_1}^x\hat\varphi_{0,\uparrow}(x_1)\hat\varphi_{l,\uparrow}(x_1)\,dx_1\Big\}.
\la{a31}\end{equation} Convergence of $\int\limits_{+\infty}^x \hat\varphi_{0,\uparrow}(x_1)
\varphi_{l,\uparrow}(x_1)\,dx_1$ follows from the fact that due to \gl{2.3}, \gl{2.4} and Lemma 5 there is
constant $C_4>0$ such that
$$\Big|\int\limits_{+\infty}^x
\hat\varphi_{0,\uparrow}(x_1)\varphi_{l,\uparrow}(x_1)\,dx_1\Big|\leqslant
C_4\int\limits_{x}^{+\infty}{{dx_1}\over\sqrt{|V(x_1)|}}
\Big(\int\limits_{x_1}^{+\infty}{{dx_2}\over\sqrt{|V(x_2)|}}\Big)^l=$$
$${C_4\over{l+1}}\Big(\int\limits_{x}^{+\infty}{{dx_2}\over
\sqrt{|V(x_2)|}}\Big)^{l+1}<+\infty.$$ Convergence of $\int\limits_{+\infty}^x\varphi_{0,\uparrow}(x_1)
\hat\varphi_{l,\uparrow}(x_1)\,dx_1$ can be proved analogously. Convergence of the integral
$\int\limits_{+\infty}^x\varphi_{0,\uparrow}(x_1)\varphi_{l,\uparrow}(x_1)\,dx_1$ is obvious. Thus the right
sides of \gl{a30} and \gl{a31} are well defined. The fact that the right sides of \gl{a30} and \gl{a31} satisfy
\gl{2.1} and \gl{2.2} for $n=l+1$ can be checked by direct application of $h$ to these sides. One must take into
account here that the Wronskian $\hat\varphi'_{0,\uparrow}(x)
\varphi_{0,\uparrow}(x)-\hat\varphi_{0,\uparrow}(x)\varphi'_{0,\uparrow}(x)\equiv2$, that follows from the
asymptotics of Lemma~8.

Now we transform the integrals in \gl{a30} and \gl{a31}.  In view of \gl{2.3} and \gl{2.4} the integrand of
$\!\!\int\limits_{+\infty}^x\!\!\!\!\hat\varphi_{0,\uparrow}(x_1)\varphi_{l,\uparrow}(x_1)\,dx_1$ reads
\begin{equation}\hat\varphi_{0,\uparrow}(x)\varphi_{l,\uparrow}(x)={1\over{2^l l!
\sqrt{V(x)-\lambda}}}\Big(\int\limits_{+\infty}^x{{dx_1}\over\sqrt{V(x_1)-
\lambda}}\Big)^l\Big[1+\Big({1\over{\xi_\uparrow(x)}}\Big)\Big],\qquad x\to+\infty. \la{a32}\end{equation} The
first term of the right side of \gl{a32} in the integral is equal to
\begin{equation}{1\over{2^l l!}}\int\limits_{+\infty}^x{{dx_1}\over
\sqrt{V(x_1)-\lambda}}\Big(\int\limits_{+\infty}^{x_1}{{dx_2}\over
\sqrt{V(x_2)-\lambda}}\Big)^l={2\over{(l+1)!}}\Big({1\over2}
\int\limits_{+\infty}^{x}{{dx_1}\over\sqrt{V(x_1)-\lambda}}\Big)^{l+1}, \la{a33}\end{equation} and the absolute
value of contribution of the second term is less than or equal to
$$C_5\int\limits_x^{+\infty}{{dx_1}\over\sqrt{|V(x_1)-\lambda|}}{1\over
{\xi_\uparrow(x_1)}}\Big(\int\limits_{x_1}^{+\infty}{{dx_2}\over\sqrt{|V(x_2)- \lambda|}}\Big)^l, $$ for some
constant $C_5>0$. From Lemma~5 the latter expression can be estimated in the following way,
$${C_5\over{\xi_\uparrow(x)}}\int\limits_x^{+\infty}{{dx_1}\over\sqrt{|V(x_1)-
\lambda|}}\Big(\int\limits_{x_1}^{+\infty}{{dx_2}\over\sqrt{|V(x_2)-
\lambda|}}\Big)^l={C_5\over{(l+1)\xi_\uparrow(x)}}\Big(\int\limits_{x}^{+\infty}
{{dx_1}\over\sqrt{|V(x_1)-\lambda|}}\Big)^{l+1}\leqslant$$
$${C_6\over{\xi_\uparrow(x)}}\Big(\int\limits_{x}^{+\infty}
{{\rm{Re}}\,\sqrt{V^*(x_1)-\lambda^*}\over{|V(x_1)-\lambda|}}
{dx_1}\Big)^{l+1}\leqslant{C_6\over{\xi_\uparrow(x)}}\Big|\int\limits_{x}^{+\infty}
{\sqrt{V^*(x_1)-\lambda^*}\over{|V(x_1)-\lambda|}} {dx_1}\Big|^{l+1}=$$
\begin{equation}{C_6\over{\xi_\uparrow(x)}}\Big|\int\limits_{x}^{+\infty}
{{dx_1}\over\sqrt{V(x_1)-\lambda}}\Big|^{l+1},\la{a33'}\end{equation}
 for some constant $C_6>0$.
Thus
\begin{equation}\int\limits_{+\infty}^x\hat\varphi_{0,\uparrow}(x_1)\varphi_{l,\uparrow}(x_1)
\,dx_1={2\over{(l+1)!}}\Big({1\over2} \int\limits_{+\infty}^{x}{{dx_1}\over\sqrt{V(x_1)-\lambda}}\Big)^{l+1}
\Big[1+\Big({1\over{\xi_\uparrow(x)}}\Big)\Big],\qquad x\to+\infty. \la{a34}\end{equation} In view of \gl{2.3},
$V(x)\in\cal K$, Lemmas~5 and 7   the following estimate for the integral
$\int\limits_{+\infty}^x\varphi_{0,\uparrow}(x_1)\varphi_{l,\uparrow}(x_1) \,dx_1$ is valid\footnote{The latter
equality in \gl{a35} is obtained with the help of the same trick as in \gl{a33'}},
$$\Big|\int\limits_{+\infty}^x\varphi_{0,\uparrow}(x_1)\varphi_{l,\uparrow}
(x_1)\,dx_1\Big|\leqslant C_7\int\limits_x^{+\infty}{{dx_1}\over\sqrt{|V(x_1)-
\lambda|}}e^{-2{\rm{Re}}\,\xi_\uparrow(x_1;\lambda)}\Big(\int\limits_{x_1}^{+\infty}
{{dx_2}\over\sqrt{|V(x_2)-\lambda|}}\Big)^l\leqslant$$
$$C_8e^{-2{\rm{Re}}\,\xi_\uparrow(x;\lambda)}\int\limits_x^{+\infty}
{{e^{-2\int\limits_{x}^{x_1}{\rm{Re}}\,\sqrt{V(x_3)-\lambda}\,dx_3}}\over
\sqrt{|V(x_1)|}}\Big(\int\limits_{x_1}^{+\infty} {{dx_2}\over\sqrt{|V(x_2)|}}\Big)^l{dx_1}\leqslant$$
$$C_8e^{-2{\rm{Re}}\,\xi_\uparrow(x;\lambda)}\int\limits_x^{+\infty}
{{e^{-C_9\int\limits_{x}^{x_1}\sqrt{|V(x_3)|}\,dx_3}}\over \sqrt{|V(x_1)|}}\Big(\int\limits_{x_1}^{+\infty}
{{dx_2}\over\sqrt{|V(x_2)|}}\Big)^l{dx_1}=$$
$$C_8e^{-2{\rm{Re}}\,\xi_\uparrow(x;\lambda)+C_9\xi_\uparrow(x)}\int\limits_x^{+\infty}
{{e^{-C_9\xi_\uparrow(x_1)}}\over \sqrt{|V(x_1)|}}\Big(\int\limits_{x_1}^{+\infty}
{{dx_2}\over\sqrt{|V(x_2)|}}\Big)^l{dx_1}=$$
$${C_8\over C_9}e^{-2{\rm{Re}}\,\xi_\uparrow(x;\lambda)+C_9\xi_\uparrow(x)}\Bigg\{
{e^{-C_9\xi_\uparrow(x)}\over{|V(x)|}}\Big(\int\limits_{x}^{+\infty} {{dx_2}\over\sqrt{|V(x_2)|}}\Big)^l+$$
$$\int\limits_x^{+\infty} e^{-C_9\xi_\uparrow(x_1)}\Big[-{l\over{|V(x_1)|^{3/2}}}
\Big(\int\limits_{x_1}^{+\infty}{{dx_2}\over\sqrt{|V(x_2)|}}\Big)^{l-1}-$$
$${{{\rm{Re}}\,V(x_1){\rm{Re}}\,V'(x_1)+{\rm{Im}}\,V(x_1){\rm{Im}}\,V'(x_1)}
\over{|V(x_1)|^3}}\Big(\int\limits_{x_1}^{+\infty}{{dx_2}\over
\sqrt{|V(x_2)|}}\Big)^{l}\Big]{dx_1}\Bigg\}\leqslant$$
$${C_8\over C_9}e^{-2{\rm{Re}}\,\xi_\uparrow(x;\lambda)}\Big[{1\over{|V(x)|}}
\Big(\int\limits_{x}^{+\infty}{{dx_2}\over\sqrt{|V(x_2)|}}\Big)^l+$$ $$C_{10}
e^{C_9\xi_\uparrow(x)}\int\limits_x^{+\infty}{{e^{-C_9\xi_\uparrow(x_1)}}\over
\sqrt{|V(x_1)|}}{1\over{\xi_\uparrow(x_1)}}\Big(\int\limits_{x_1}^{+\infty}
{{dx_2}\over\sqrt{|V(x_2)|}}\Big)^{l}{dx_1}\Big]=$$
$$O\Big(e^{-2\xi_\uparrow(x;\lambda)}{1\over{V(x)}}\Big(\int\limits_{x}^{+\infty}
{{dx_2}\over\sqrt{|V(x_2)|}}\Big)^l\Big)+O\Big(e^{-2\xi_\uparrow(x;\lambda)}
{1\over{\xi_\uparrow(x)}}\Big(\int\limits_{x}^{+\infty}{{dx_2}\over\sqrt{|V(x_2)|}} \Big)^{l+1}\Big)=$$
\begin{equation}O\Big({e^{-2\xi_\uparrow(x;\lambda)}\over{\xi_\uparrow(x)}}\Big(
\int\limits_{x}^{+\infty}{{dx_2}\over\sqrt{|V(x_2)-\lambda|}}
\Big)^{l+1}\Big)=O\Big({e^{-2\xi_\uparrow(x;\lambda)}\over{\xi_\uparrow(x)}}\Big(
\int\limits_{x}^{+\infty}{{dx_2}\over\sqrt{V(x_2)-\lambda}} \Big)^{l+1}\Big),\la{a35}\end{equation} for some
positive constants~$C_7$,~\dots $C_{10}$. The asymptotics \gl{2.3} and \gl{2.5} for $n=l+1$ follow from
\gl{a30}, \gl{a34}, \gl{a35}, from \gl{2.3} and \gl{2.4} with $n=0$ and from \gl{a13}, \gl{a15} and Corollary~1.

The integral $\int\limits_{+\infty}^x\varphi_{0,\uparrow}(x_1)\hat\varphi_{l,\uparrow}(x_1) \,dx_1$ can be
calculated in the same way and the result is
\begin{equation}\int\limits_{+\infty}^x\varphi_{0,\uparrow}(x_1)\hat\varphi_{l,\uparrow}(x_1)
\,dx_1={{-2}\over{(l+1)!}}\Big(-{1\over2} \int\limits_{+\infty}^{x}{{dx_1}\over\sqrt{V(x_1)-\lambda}}\Big)^{l+1}
\Big[1+O\Big({1\over{\xi_\uparrow(x)}}\Big)\Big],\quad x\to+\infty. \la{a36}\end{equation} One can also obtain
the estimate,
\begin{equation}\int\limits_{R_1}^x\hat\varphi_{0,\uparrow}(x_1)
\hat\varphi_{l,\uparrow}(x_1)\,dx_1=O\bigg({e^{2\xi_\uparrow(x;\lambda)}\over{\xi_\uparrow(x)}}
\Big(\int\limits_{+\infty}^{x}{{dx_1}\over\sqrt{V(x_1)-\lambda}}\Big)^{l+1} \bigg),\qquad
x\to+\infty.\la{a37}\end{equation} Then in view of \gl{2.3}, \gl{2.4} for $n=0$ and \gl{a31}, \gl{a36}, \gl{a37}
the asymptotics \gl{2.4} turns out to be valid for $n=l+1$. For the integral
$\int\limits_{R_1}^x\hat\varphi_{0,\uparrow}(x_1)\hat\varphi_{l,\uparrow}(x_1)\,dx_1$ the following estimate can
be derived for some positive constants $C_{11}$,~\dots, $C_{15}$, $\xi_0$,
$$\Big|\int\limits_{R_1}^x\hat\varphi_{0,\uparrow}(x_1)\hat\varphi_{l,\uparrow}
(x_1)\,dx_1\Big|\leqslant C_{11}\int\limits_{R_1}^x{e^{2{\rm{Re}}\,\xi_\uparrow(x_1;\lambda)}\over\sqrt{|V(x_1)-
\lambda|}}\Big(\int\limits_{x_1}^{+\infty} {{dx_2}\over\sqrt{|V(x_2)-\lambda|}}\Big)^l{dx_1}\leqslant$$
$$C_{12}e^{2{\rm{Re}}\,\xi_\uparrow(x;\lambda)}\int\limits_{R_0}^x
{{e^{-C_{13}(\xi_\uparrow(x)-\xi_\uparrow(x_1))}}\over \sqrt{|V(x_1)|}}\Big(\int\limits_{x_1}^{+\infty}
{{dx_2}\over\sqrt{|V(x_2)|}}\Big)^l{dx_1}=$$
$${C_{12}\over C_{13}}e^{2{\rm{Re}}\,\xi_\uparrow(x;\lambda)-C_{13}\xi_\uparrow(x)}\Bigg\{
{e^{C_{13}\xi_\uparrow(x_1)}\over{|V(x_1)|}}\Big(\int\limits_{x_1}^{+\infty}
{{dx_2}\over\sqrt{|V(x_2)|}}\Big)^l\Bigg|_{R_0}^{x}-$$
$$\int\limits_{R_0}^x e^{C_{13}\xi_\uparrow(x_1)}\Big[-{l\over{|V(x_1)|^{3/2}}}
\Big(\int\limits_{x_1}^{+\infty}{{dx_2}\over\sqrt{|V(x_2)|}}\Big)^{l-1}$$
$$-{{{\rm{Re}}\,V(x_1){\rm{Re}}\,V'(x_1)+{\rm{Im}}\,V(x_1){\rm{Im}}\,V'(x_1)}
\over{|V(x_1)|^3}}\Big(\int\limits_{x_1}^{+\infty}{{dx_2}\over
\sqrt{|V(x_2)|}}\Big)^{l}\Big]{dx_1}\Bigg\}\leqslant$$
$${C_{12}\over C_{13}}e^{2{\rm{Re}}\,\xi_\uparrow(x;\lambda)}\Big[{1\over{|V(x)|}}
\Big(\int\limits_{x}^{+\infty}{{dx_2}\over\sqrt{|V(x_2)|}}\Big)^l+$$ $$C_{14}
e^{-C_{13}\xi_\uparrow(x)}\int\limits_{R_0}^x{{e^{C_{13}\xi_\uparrow(x_1)}}\over
\sqrt{|V(x_1)|}}\Big(\int\limits_{x_1}^{+\infty}
{{dx_2}\over\sqrt{|V(x_2)|}}\Big)^{l}{{dx_1}\over{\xi_0+\xi_\uparrow(x_1)}}\Big]\leqslant$$
$${C_{12}\over
C_{13}}e^{2{\rm{Re}}\,\xi_\uparrow(x;\lambda)}\Big[{C_{15}\over{\xi_0+\xi_\uparrow(x)}}
\Big(\int\limits_{x}^{+\infty}{{dx_2}\over\sqrt{|V(x_2)|}}\Big)^{l+1}+$$ \begin{equation}C_{14}
e^{-C_{13}\xi_\uparrow(x)}\int\limits_{R_0}^x{{e^{C_{13}\xi_\uparrow(x_1)}}\over
\sqrt{|V(x_1)|}}\Big(\int\limits_{x_1}^{+\infty}
{{dx_2}\over\sqrt{|V(x_2)|}}\Big)^{l}{{dx_1}\over{\xi_0+\xi_\uparrow(x_1)}}\Big], \la{a38}\end{equation} with
the help of \gl{2.4} for $n=0$ and $n=l$, $V(x)\in\cal K$ and Lemmas~5 and~7. Let us show that
$$e^{-C_{13}\xi_\uparrow(x)}\int\limits_{R_0}^x{{e^{C_{13}\xi_\uparrow(x_1)}}\over
\sqrt{|V(x_1)|}}\Big(\int\limits_{x_1}^{+\infty}{{dx_2}\over\sqrt{|V(x_2)|}}
\Big)^{l}{{dx_1}\over{\xi_0+\xi_\uparrow(x_1)}}=$$ \begin{equation} o\Big({1\over{\xi_\uparrow(x)}}
\Big(\int\limits_{x}^{+\infty}{{dx_2}\over\sqrt{|V(x_2)|}}\Big)^{l+1}\Big), \qquad x\to+\infty.
\la{a39}\end{equation} Then using \gl{a38} and the fact that in accordance to Lemma~5
$$\int\limits_{x}^{+\infty}{{dx_2}\over\sqrt{|V(x_2)|}}=O\Big(
\int\limits_{x}^{+\infty}{{{\rm{Re}}\,\sqrt{V^*(x_2)-\lambda^*}} \over{|V(x_2)-\lambda|}}dx_2\Big)=$$
$$O\Big(\int\limits_{x}^{+\infty} {{\sqrt{V^*(x_2)-\lambda^*}}\over{|V(x_2)-\lambda|}}dx_2\Big)=O\Big(
\int\limits_{x}^{+\infty}{{dx_2}\over\sqrt{V(x_2)-\lambda}}\Big), \qquad x\to+\infty,$$ the required estimate
\gl{a37} would be proved.

Performing the change of variable $\xi=\xi_\uparrow(x_2)$, we get
$$e^{-C_{13}\xi_\uparrow(x)}\int\limits_{R_0}^x
{{e^{C_{13}\xi_\uparrow(x_1)}}\over\sqrt{|V(x_1)|}}\Big(\int\limits_{x_1}^{+\infty}
{{dx_2}\over\sqrt{|V(x_2)|}}\Big)^l{{dx_1}\over{\xi_0+\xi_\uparrow(x_1)}}=$$
$$e^{-C_{13}\xi_\uparrow(x)}\Big(\int\limits_0^{\xi_\uparrow(x)/2}+\int\limits_{\xi_\uparrow(x)/2}^{
\xi_\uparrow(x)}\Big){{e^{C_{13}\xi}}\over{|V(x_1(\xi))|}}\Big(\int\limits_{
x_1(\xi)}^{+\infty}{{dx_2}\over\sqrt{|V(x_2)|}}\Big)^l{{d\xi} \over{\xi_0+\xi}}\leqslant$$
$$e^{-C_{13}\xi_\uparrow(x)}\Bigg\{{1\over{\xi_0}}e^{C_{13}\xi_\uparrow(x)/2}
\int\limits_0^{\xi_\uparrow(x)/2}\Big(\int\limits_{x_1(\xi)}^{+\infty}
{{dx_2}\over\sqrt{|V(x_2)|}}\Big)^l{{d\xi}\over{|V(x_1(\xi))|}}+$$
$${1\over{\xi_0+\xi_\uparrow(x)/2}}\int\limits_{\xi_\uparrow(x)/2}^{\xi_\uparrow(x)}
{{e^{C_{13}\xi}}\over{|V(x_1(\xi))|}}\Big(\int\limits_{ x_1(\xi)}^{+\infty}{{dx_2}\over\sqrt{|V(x_2)|}}\Big)^l
d\xi\Bigg\}\leqslant$$
$${e^{-C_{13}\xi_\uparrow(x)/2}\over{\xi_0}}\!\!
\int\limits_{R_0}^{+\infty}\!\!\Big(\!\!\int\limits_{x_1}^{+\infty}\!\!
{{dx_2}\over\sqrt{|V(x_2)|}}\Big)^l{{dx_1}\over\sqrt{|V(x_1)|}}+
{{2e^{-C_{13}\xi_\uparrow(x)}}\over{\xi_\uparrow(x)}}\!\!\int\limits_{R_0}^{x}\!\!
{{e^{C_{13}\xi_\uparrow(x_1)}}\over\sqrt{|V(x_1)|}}\Big(\!\!\int\limits_{
x_1}^{+\infty}\!\!{{dx_2}\over\sqrt{|V(x_2)|}}\Big)^l dx_1\!=$$
\begin{equation}{e^{-C_{13}\xi_\uparrow(x)/2}\over{(l+1)\xi_0}}
\Big(\int\limits_{R_0}^{+\infty}{{dx_1}\over\sqrt{|V(x_1)|}}\Big)^{l+1}+
{{2e^{-C_{13}\xi_\uparrow(x)}}\over{\xi_\uparrow(x)}}\int\limits_{R_0}^{x}
{{e^{C_{13}\xi_\uparrow(x_1)}}\over\sqrt{|V(x_1)|}}\Big(\int\limits_{
x_1}^{+\infty}{{dx_2}\over\sqrt{|V(x_2)|}}\Big)^l dx_1.\la{a41}\end{equation} It follows from \gl{a41} and from
the estimate of $\int\limits_{R_0}^x
{{e^{C_{13}\xi_\uparrow(x_1)}}\over\sqrt{|V(x_1)|}}\Big(\int\limits_{x_1}^{+\infty}
{{dx_2}\over\sqrt{|V(x_2)|}}\Big)^l{dx_1}$ contained in \gl{a38}, that
$$e^{-C_{13}\xi_\uparrow(x)}\int\limits_{R_0}^x
{{e^{C_{13}\xi_\uparrow(x_1)}}\over\sqrt{|V(x_1)|}}\Big(\int\limits_{x_1}^{+\infty}
{{dx_2}\over\sqrt{|V(x_2)|}}\Big)^l{{dx_1}\over{\xi_0+\xi_\uparrow(x_1)}}=$$
\begin{equation}O\big(e^{-C_{13}\xi_\uparrow(x)/2}\big)+o\Big({1\over{\xi_\uparrow(x)}}
\Big(\int\limits_{x}^{+\infty}{{dx_2}\over\sqrt{|V(x_2)|}}\Big)^{l+1}\Big). \la{a42}\end{equation} As well it
follows from \gl{a11} and \gl{a20} that
$$O\big(e^{-C_{13}\xi_\uparrow(x)/2}\big)=
o\big(\xi_\uparrow^{l-(l+1)\gamma}(x)\big)= o\Big({{\xi_\uparrow^{l}(x)}\over{|V(x)|^{l+1}}}\Big)=$$
\begin{equation}
o\Big({1\over{\xi_\uparrow(x)}} \Big(\int\limits_{x}^{+\infty}{{dx_2}\over\sqrt{|V(x_2)|}}
\Big)^{l+1}\Big),\qquad x\to+\infty.\la{a40}\end{equation} Then the estimate \gl{a39} is derived from \gl{a40}
and \gl{a42}. Thus, \gl{2.4} is valid for $n=l+1$.

Finally let us show that functions  $\varphi_{n,\uparrow}(x)$ ($\hat\varphi_{n,\uparrow}(x)$) for any $n$ are
normalizable (non-normalizable) at $+\infty$. For this purpose it is sufficient  to prove that the leading term
of \gl{2.3} (\gl{2.4}) is normalizable (non-normalizable) at $+\infty$. Normalizability of the leading term of
\gl{2.3} is owed to the fact that, because of Lemma~5, the following estimates take place,
$${1\over\sqrt{|V(x)-\lambda|}}e^{-2{\rm{Re}}\,\xi_\uparrow(x;\lambda)}\Big|
\int\limits_{+\infty}^x{{dx_1}\over\sqrt{V(x_1)-\lambda}}\Big|^{2n}\leqslant$$
$${C_2^{(2n+1)/4}\over\sqrt{|V(x)|}}e^{-2C_3(\xi_\uparrow(x)-\xi_\uparrow(R_1))}\Big(
\int\limits_x^{+\infty}{{dx_1}\over\sqrt{|V(x_1)|}}\Big)^{2n}\leqslant$$
$$C_2^{(2n+1)/4}{{\xi'_\uparrow(x)}\over{|V(x)|}}e^{-2C_3(\xi_\uparrow(x)-
\xi_\uparrow(R_1))}\Big( \int\limits_{R_0}^{+\infty}{{dx_1}\over\sqrt{|V(x_1)|}}\Big)^{2n}\leqslant$$
$${C_2^{(2n+1)/4}\over\varepsilon}\Big(\int\limits_{R_0}^{+\infty}{{dx_1}
\over\sqrt{|V(x_1)|}}\Big)^{2n}\xi'_\uparrow(x)e^{-2C_3(\xi_\uparrow(x)-\xi_\uparrow(R_1))}, \qquad x\geqslant
R_1,$$ where the latter expression is  normalizable at $+\infty$. Non-normalizability of the leading term
\gl{2.4} follows from the fact that in view of Lemma 5 and \gl{a11}, \gl{a20} the estimates are valid for some
constant $C_{16}>0$,
$${1\over\sqrt{|V(x)-\lambda|}}e^{2{\rm{Re}}\,\xi_\uparrow(x;\lambda)}\Big|
\int\limits_{+\infty}^x{{dx_1}\over\sqrt{V(x_1)-\lambda}}\Big|^{2n}\geqslant$$
$${{\root 4 \of C_1}\over\sqrt{|V(x)|}}e^{2C_3(\xi_\uparrow(x)-\xi_\uparrow(R_1))}\Big(
\int\limits_x^{+\infty}{{{\rm{Re}\,}\sqrt{V^*(x_1)-\lambda^*}}\over{|V(x_1) -\lambda|}}dx_1\Big)^{2n}\geqslant$$
$${{C_1^{(4n+1)/4}C_3^{2n}}\over{|V(x)|}}\xi'_\uparrow(x)e^{2C_3(\xi_\uparrow(x)-
\xi_\uparrow(R_1) )}\Big(\int\limits_x^{+\infty}{{dx_1}\over\sqrt{|V(x_1)|}}\Big)^{2n}\geqslant$$
\begin{equation}C_{16}{{\xi'_\uparrow(x)\xi_\uparrow^{2n}(x)}\over{|V(x)|^{2n+1}}}
e^{2C_3\xi_\uparrow(x)}\geqslant {C_{16}\over
C_0^{2n+1}}{{\xi'_\uparrow(x)\xi_\uparrow^{2n}(x)}\over{(\xi_0+\xi_\uparrow(x))^{
(2n+1)\gamma}}}e^{2C_3\xi_\uparrow(x)},\qquad x\geqslant R_1,\la{a42'}\end{equation} where the latter expression
is non-normalizable at $+\infty$.

Let us now  consider the case $\int\limits_{R_0}^{+\infty}dx_1/\sqrt{ |V(x_1)|}=+\infty$ and prove that
$\varphi_{l+1,\uparrow}(x)$ and $\hat \varphi_{l+1,\uparrow}(x)$ can be written in the form
\begin{equation}\varphi_{l+1,\uparrow}(x)=-{1\over2}\Big\{\hat\varphi_{0,\uparrow}(x)\int
\limits_{+\infty}^x\varphi_{0,\uparrow}(x_1)\varphi_{l,\uparrow}(x_1)\,dx_1-\varphi_{0,\uparrow}(x)
\int\limits_{R_1}^x\hat\varphi_{0,\uparrow}(x_1)\varphi_{l,\uparrow}(x_1)\,dx_1\Big\}, \la{a43}\end{equation}
\begin{equation}\hat\varphi_{l+1,\uparrow}(x)=-{1\over2}\Big\{\hat\varphi_{0,\uparrow}(x)\int
\limits_{R_1}^x\varphi_{0,\uparrow}(x_1)\hat\varphi_{l,\uparrow}(x_1)\,dx_1-\varphi_{0,\uparrow}(x)
\int\limits_{R_1}^x\hat\varphi_{0,\uparrow}(x_1)\hat\varphi_{l,\uparrow}(x_1)\,dx_1\Big\}.
\la{a44}\end{equation} Convergence of $\int\limits_{+\infty}^x\varphi_{0,\uparrow}(x_1)\varphi_{l,\uparrow}(x_1)
\,dx_1$ follows from the fact that  $V(x)\in\cal K$ and  in view of Lemma~5, and \gl{2.6} for $n=0$ and $n=l$
there are positive constants $C_3$, $C_{17}$, $\varepsilon$ such that
$$\Big|\int\limits_{+\infty}^x\varphi_{0,\uparrow}(x_1)\varphi_{l,\uparrow}(x_1)
\,dx_1\Big|\leqslant C_{17}e^{-2{\rm{Re}}\,\xi_\uparrow(x;\lambda)+2C_3\xi_\uparrow(x)}
\int\limits_x^{+\infty}{e^{-2C_3\xi_\uparrow(x_1)}\over\sqrt{|V(x_1)|}}\Big(
\int\limits_{R_1}^{x_1}{{dx_2}\over\sqrt{|V(x_2)|}}\Big)^l dx_1\leqslant$$
$${C_{17}\over\varepsilon^l}e^{-2{\rm{Re}}\,\xi_\uparrow(x;\lambda)+2C_3\xi_\uparrow(x)}
\int\limits_x^{+\infty}{{\xi_\uparrow^l(x_1)}\over\sqrt{|V(x_1)|}}e^{-2C_3\xi_\uparrow(x_1)} dx_1\leqslant$$
$${C_{17}\over\varepsilon^{l+1}}e^{-2{\rm{Re}}\,\xi_\uparrow(x;\lambda)+2C_3\xi_\uparrow(x)}
\int\limits_x^{+\infty}\xi_\uparrow^l(x_1)\xi'_\uparrow(x_1)e^{-2C_3\xi_\uparrow(x_1)} dx_1=$$
$${C_{17}\over\varepsilon^{l+1}}e^{-2{\rm{Re}}\,\xi_\uparrow(x;\lambda)+2C_3\xi_\uparrow(x)}
\int\limits_{\xi_\uparrow(x)}^{+\infty}\xi^le^{-2C_3\xi}d\xi<+\infty.$$ Let us now find the asymptotics of
integrals, contained in \gl{a43} and \gl{a44}. Due to \gl{2.6} and \gl{2.7} the integrand of
$\int\limits_{R_1}^x\hat\varphi_{0,\uparrow}(x_1) \varphi_{l,\uparrow}(x_1)\,dx_1$ reads
\begin{equation}\hat\varphi_{0,\uparrow}(x)\varphi_{l,\uparrow}(x)={1\over{2^l l!
\sqrt{V(x)-\lambda}}}\Big(\int\limits_{R_1}^x{{dx_1}\over\sqrt{V(x_1)-
\lambda}}\Big)^l\Big[1+O\Big({{\ln\eta_\uparrow(x)}\over{\eta_\uparrow(x)}}\Big)\Big], \quad x\to+\infty.
\la{a45}\end{equation} The first term of right side of \gl{a45} contributes into the integral as follows,
\begin{equation}{1\over{2^l l!}}\int\limits_{R_1}^x\Big(\int\limits_{R_1
}^{x_1}{{dx_2}\over\sqrt{V(x_2)-\lambda}}\Big)^l{{dx_1}\over\sqrt{V(x_1)
-\lambda}}={2\over{(l+1)!}}\Big({1\over2} \int\limits_{R_1}^{x}{{dx_1}\over\sqrt{V(x_1)-\lambda}}\Big)^{l+1},
\la{a46}\end{equation} and the  absolute value of contribution of the second term is less than or equal to
$$C_{18}\int\limits_{R_1}^x\Big(\int\limits_{R_1}^{x_1}{{dx_2}\over
\sqrt{|V(x_2)-\lambda|}}\Big)^l{{\ln(2+\eta_\uparrow(x_1))}\over{2+
\eta_\uparrow(x_1)}}{{dx_1}\over\sqrt{|V(x_1)-\lambda|}},$$ for a constant $C_{18}>0$.
 In view of Lemma 5 the latter expression is less than or equal to
$$C_{19}\int\limits_{R_0}^x\Big(\int\limits_{R_0}^{x_1}{{dx_2}\over\sqrt{
|V(x_2)|}}\Big)^l{{\ln(2+\eta_\uparrow(x_1))}\over{2+\eta_\uparrow(x_1)}}
{{dx_1}\over\sqrt{|V(x_1)|}}\leqslant$$
$$C_{19}\int\limits_{R_0}^x\eta'_\uparrow(x_1)(2+\eta_\uparrow(x_1))^{l-1}
\ln(2+\eta_\uparrow(x_1))\,dx_1\leqslant$$
$$C_{19}(2+\eta_\uparrow(x))^l\ln(2+\eta_\uparrow(x))=O\Big[{{\ln\eta_\uparrow
(x)}\over{\eta_\uparrow(x)}}\Big(\int\limits_{R_1}^x {{dx_1}\over\sqrt{|V(x_1)|}}\Big)^{l+1}\Big]=$$
$$O\Big[{{\ln\eta_\uparrow
(x)}\over{\eta_\uparrow(x)}}\Big(\int\limits_{R_1}^x{{{\rm{Re}}\,\sqrt{V^*
(x_1)-\lambda^*}}\over{|V(x_1)-\lambda|}}\,dx_1\Big)^{l+1}\Big]=$$ \begin{equation}O\Big[{{\ln
\eta_\uparrow(x)}\over{\eta_\uparrow(x)}}\Big(\int\limits_{R_1}^x
{{dx_1}\over\sqrt{V(x_1)-\lambda}}\Big)^{l+1}\Big],\qquad x\to+\infty,\la{a47'} \end{equation} for a constant
$C_{19}>0$. Thus,
\begin{equation}\!\int\limits_{R_1}^x\!\hat\varphi_{0,\uparrow}(x_1)\varphi_{l,\uparrow}(x_1)
\,dx_1={2\over{(l+1)!}}\Big({1\over2} \int\limits_{R_1}^{x}{{dx_1}\over\sqrt{V(x_1)-\lambda}}\Big)^{l+1}
\Big[1+O\Big({{\ln\eta_\uparrow(x)}\over{\eta_\uparrow(x)}}\Big)\Big],\quad x\to+\infty. \la{a48}\end{equation}

In the case $l=0$ one may  write \gl{a45}, using \gl{a123}, \gl{a14} and Lemma~6 in the form
\begin{equation}\hat\varphi_{0,\uparrow}(x)\varphi_{0,\uparrow}(x)={1\over\sqrt{V(x)-\lambda}}
\Big[1+O\Big({1\over{\xi_\uparrow(x)}}\Big)\Big],\qquad x\to+\infty.\la{a47}
\end{equation}
Respectively, due to Lemma~5 and for $V(x)\in\cal K$  the contribution of the second term of \gl{a47} to the
integral $\int\limits_{R_1}^x\hat\varphi_{0,\uparrow}(x_1) \varphi_{0,\uparrow}(x_1)\,dx_1$ is less than or
equal to,
$$C_{20}\int\limits_{R_1}^x{{dx_1}\over{\sqrt{|V(x_1)-\lambda|}(\xi_0+
\xi_\uparrow(x_1))}}\leqslant C_{20}\root4\of{C_2}\int\limits_{R_0}^x{{dx_1}\over{\sqrt{|V(x_1)|}
(\xi_0+\int_{R_0}^{x_1}\sqrt{|V(x_2)|}\,dx_2)}}\leqslant$$
$$C_{20}\root4\of{C_2}\int\limits_{R_0}^x{{\eta'_\uparrow(x_1)\,dx_1}\over
{\xi_0+\varepsilon\eta_\uparrow(x_1)}}={{C_{20}\root4\of{C_2}}
{1\over\varepsilon}}\ln[1+\varepsilon\eta_\uparrow(x)/\xi_0]=O(\ln\eta_\uparrow(x))=$$
$$O\Big({{\ln\eta_\uparrow(x)}\over{\eta_\uparrow(x)}}\int\limits_{R_1}^x
{{dx_1}\over\sqrt{V(x_1)-\lambda}}\Big),\qquad x\to+\infty,$$ for constants $C_{20}>0$ and $\xi_0>0$.

In view of \gl{2.6}, \gl{a29'}, Lemma~5 and $V(x)\in\cal K$,  the following estimate  holds for the integral
$\int\limits_{+\infty}^x\varphi_{0,\uparrow}(x_1)\varphi_{l,\uparrow}(x_1) \,dx_1$, \footnote{In \gl{a49} and
\gl{a51} the estimate $\int_{R_1}^{x}dx_1/\sqrt{|V(x_1)|}=O(\int_{R_1}^{x}dx_1/\sqrt{V(x_1)- \lambda})$ is used.
Derivation of this estimate is contained in \gl{a47'}.}
$$\Big|\int\limits_{+\infty}^x\varphi_{0,\uparrow}(x_1)\varphi_{l,\uparrow}(x_1)\,dx_1\Big|
\leqslant C_{20}\int\limits_x^{+\infty}\Big(\int\limits_{R_1}^{x_1}{{dx_2}\over
\sqrt{|V(x_2)-\lambda|}}\Big)^le^{-2{\rm{Re}}\,\xi_\uparrow(x_1;\lambda)}{{dx_1}\over
\sqrt{|V(x_1)-\lambda|}}\leqslant$$
$$C_{21}e^{-2{\rm{Re}}\,\xi_\uparrow(x;\lambda)+C_{22}\xi_\uparrow(x)}\int\limits_x^{
+\infty}\Big(\int\limits_{R_1}^{x_1}{{dx_2}\over
\sqrt{|V(x_2)|}}\Big)^l{{e^{-C_{22}\xi_\uparrow(x_1)}\,dx_1}\over \sqrt{|V(x_1)|}}=$$
$$-{C_{21}\over
C_{22}}e^{-2{\rm{Re}}\,\xi_\uparrow(x;\lambda)+C_{22}\xi_\uparrow(x)}\Bigg\{
{{e^{-C_{22}\xi_\uparrow(x_1)}}\over {|V(x_1)|}}\eta_\uparrow^l(x_1)\Big|_{x}^{+\infty}-$$ $$\int\limits_x^{
+\infty}e^{-C_{22}\xi_\uparrow(x_1)}\Big[{{l\eta_\uparrow^{l-1}(x_1)}\over{|V(x_1)|^{3/2}}}
-{{{\rm{Re}}\,V(x_1){\rm{Re}}\,V'(x_1)+{\rm{Im}}\,V(x_1){\rm{Im}}\,V'(x_1)}
\over{|V(x_1)|^3}}\eta_\uparrow^l(x_1)\Big]dx_1\Bigg\}\leqslant$$
$${C_{21}\over C_{22}}e^{-2{\rm{Re}}\,\xi_\uparrow(x;\lambda)}\Big[{{\eta_\uparrow^l(x)}
\over{|V(x)|}}+e^{C_{22}\xi_\uparrow(x)}\int\limits_x^{
+\infty}{e^{-C_{22}\xi_\uparrow(x_1)}\over\sqrt{|V(x_1)|}}\Big(
{l\over\varepsilon}\eta_\uparrow^{l-1}(x_1)+C_{23}
{\eta_\uparrow^l(x_1)\over{\xi_\uparrow(x_1)}}\Big)dx_1\Big]\leqslant$$
$${C_{21}\over{\varepsilon
C_{22}}}e^{-2{\rm{Re}}\,\xi_\uparrow(x;\lambda)}\Big[\eta_\uparrow^l(x)+
e^{C_{22}\xi_\uparrow(x)}\int\limits_x^{
+\infty}{e^{-C_{22}\xi_\uparrow(x_1)}\over\sqrt{|V(x_1)|}}\eta_\uparrow^{l-1}(x_1)( l+C_{23})\,dx_1\Big]=$$
$$O[e^{-2{\rm{Re}}\,\xi_\uparrow(x;\lambda)}\eta_\uparrow^l(x)]+O\Big[e^{-2{\rm{Re}}\,
\xi_\uparrow(x;\lambda)+C_{22}\xi_\uparrow(x)}{1\over\eta_\uparrow(x)}\int\limits_x^{
+\infty}{e^{-C_{22}\xi_\uparrow(x_1)}\over\sqrt{|V(x_1)|}}\eta_\uparrow^{l}(x_1) \,dx_1\Big]=$$
\begin{equation}O\Big[e^{-2{\rm{Re}}\,\xi_\uparrow(x;\lambda)}\Big(\int
\limits_{R_1}^{x}{{dx_2}\over\sqrt{|V(x_2)|}}\Big)^l
\Big]=O\Big[e^{-2{\rm{Re}}\,\xi_\uparrow(x;\lambda)}\Big(\int
\limits_{R_1}^{x}{{dx_2}\over\sqrt{V(x_2)-\lambda}}\Big)^l \Big],\qquad x\to+\infty,\la{a49}\end{equation} for
positive constants~$C_{20}$,~\dots, $C_{23}$ The asymptotics \gl{2.6} and \gl{2.8} for $n=l+1$ is derived from
\gl{a43}, \gl{a48}, \gl{a49} from \gl{2.6} and \gl{2.7} for $n=0$ and from \gl{a13}, \gl{a15}, \gl{a29'} and
corollary~1.

The integral $\int\limits_{R_1}^x\varphi_{0,\uparrow}(x_1)\hat\varphi_{l,\uparrow}(x_1) \,dx_1$ can be
calculated in the same way as the following one, $\int\limits_{R_1}^x\hat\varphi_{0,\uparrow}(x_1)
\varphi_{l,\uparrow}(x_1) \,dx_1$ and the result is
\begin{equation}\int\limits_{R_1}^x\varphi_{0,\uparrow}(x_1)\hat\varphi_{l,\uparrow}(x_1)
\,dx_1=-{2\over{(l+1)!}}\Big(-{1\over2} \int\limits_{R_1}^{x}{{dx_1}\over\sqrt{V(x_1)-\lambda}}\Big)^{l+1}
\Big[1+O\Big({{\ln\eta_\uparrow(x)}\over{\eta_\uparrow(x)}}\Big)\Big],\,\,\, x\to+\infty.\la{a50}\end{equation}
For the integral $\int\limits_{R_1}^x\hat\varphi_{0,\uparrow}(x_1)\hat\varphi_{l,\uparrow}(x_1) \,dx_1$, due to
\gl{2.7}, \gl{a29'}, Lemma~5 and $V(x)\in\cal K$, the estimate  takes place,
$$\Big|\int\limits_{R_1}^x\hat\varphi_{0,\uparrow}(x_1)\hat\varphi_{l,\uparrow}(x_1)\,dx_1
\Big|\leqslant C_{24}\int\limits_{R_1}^x\Big(\int\limits_{R_1}^{x_1}{{dx_2}\over
\sqrt{|V(x_2)-\lambda|}}\Big)^l{{e^{2{\rm{Re}}\,\xi_\uparrow(x_1;\lambda)}\,dx_1}\over
\sqrt{|V(x_1)-\lambda|}}\leqslant$$
$$C_{25}e^{2{\rm{Re}}\,\xi_\uparrow(x;\lambda)-C_{26}\xi_\uparrow(x)}\int\limits_{R_0}^x
\Big(\int\limits_{R_0}^{x_1}{{dx_2}\over \sqrt{|V(x_2)|}}\Big)^l{{e^{C_{26}\xi_\uparrow(x_1)}\,dx_1}\over
\sqrt{|V(x_1)|}}=$$
$${C_{25}\over\varepsilon}e^{2{\rm{Re}}\,\xi_\uparrow(x;\lambda)-C_{26}\xi_\uparrow(x)}
\eta_\uparrow^l(x)\int\limits_{R_0}^x\xi'_\uparrow(x_1)e^{C_{26}\xi_\uparrow(x_1)}\,dx_1\leqslant
{C_{25}\over{\varepsilon C_{26}}}e^{2{\rm{Re}}\,\xi_\uparrow(x;\lambda)}\eta_\uparrow^l(x)=$$
\begin{equation}O\Big[e^{2{\rm{Re}}\,\xi_\uparrow(x;\lambda)}\Big(\int
\limits_{R_1}^{x}{{dx_2}\over\sqrt{|V(x_2)|}}\Big)^l
\Big]=O\Big[e^{2{\rm{Re}}\,\xi_\uparrow(x;\lambda)}\Big(\int
\limits_{R_1}^{x}{{dx_2}\over\sqrt{V(x_2)-\lambda}}\Big)^l \Big],\qquad x\to+\infty,\la{a51}\end{equation} for
positive constants~$C_{24}$,\dots, $C_{27}$. The asymptotics \gl{2.7} for $n=l+1$ follows from \gl{a44},
\gl{a50}, \gl{a51} as well as  from \gl{2.6} and \gl{2.7} for $n=0$.

Finally let us  check that $\varphi_{n,\uparrow}(x)$ ($\hat\varphi_{n,\uparrow}(x)$) for any $n$ is normalizable
(non-normalizable) at $+\infty$. For this purpose it is sufficient  to examine that the leading term of the
right side \gl{2.6} (\gl{2.7}) is normalizable (non-normalizable) at $+\infty$. Normalizability of the leading
term of \gl{2.6} follows from the fact that due to $V(x)\in\cal K$ and Lemma~5 the following estimate is valid,
$${1\over\sqrt{|V(x)-\lambda|}}e^{-2{\rm{Re}}\,\xi_\uparrow(x;\lambda)}\Big|
\int\limits_{R_1}^x{{dx_1}\over\sqrt{V(x_1)-\lambda}}\Big|^{2n}\leqslant
{C_2^{(2n+1)/4}\over\sqrt{|V(x)|}}e^{-2C_3(\xi_\uparrow(x)-\xi_\uparrow(R_1))}\times$$ $$\Big(
\int\limits_{R_0}^x{{dx_1}\over\sqrt{|V(x_1)|}}\Big)^{2n}\leqslant
{{C_2^{(2n+1)/4}\xi'_\uparrow(x)}\over{\varepsilon^{2n}|V(x)|}}
e^{-2C_3(\xi_\uparrow(x)-\xi_\uparrow(R_1))}\xi^{2n}_\uparrow(x)\leqslant$$ $${{C_2^{(2n+1)/4}}\over
{\varepsilon^{2n+1}}}\xi^{2n}_\uparrow(x)\xi'_\uparrow(x)e^{-2C_3(\xi_\uparrow(x)-\xi_\uparrow(R_1))}, \qquad
x\geqslant R_1,$$ the right side of which is obviously normalizable at $+\infty$. Non-normalizability of the
leading  term in \gl{2.7} follows from the fact that in view of \gl{a20}, Lemma 5 and with the help of the trick
in \gl{a42'} the following estimate holds:
$${1\over\sqrt{|V(x)-\lambda|}}e^{2{\rm{Re}}\,\xi_\uparrow(x;\lambda)}\Big|
\int\limits_{R_1}^x{{dx_1}\over\sqrt{V(x_1)-\lambda}}\Big|^{2n}\geqslant
{{C_1^{(4n+1)/4}C_3^{2n}}\over\sqrt{|V(x)|}}{\xi'_\uparrow(x)\over\sqrt{|V(x)|}}
e^{2C_3(\xi_\uparrow(x)-\xi_\uparrow(R_1))}\times$$ $$\Big(\int\limits_{R_1}^x{{dx_1}\over\sqrt
{|V(x_1)|}}\Big)^{2n}\!\geqslant\!
{{C_1^{(4n+1)/4}C_3^{2n}}\over{C_0}}{{\xi'_\uparrow(x)}\over{(\xi_0+\xi_\uparrow(x))^\gamma}}
e^{2C_3(\xi_\uparrow(x)-\xi_\uparrow(R_1))}\Big(\int\limits_{R_1}^x{{dx_1}\over
\sqrt{|V(x_1)|}}\Big)^{2n}\!\!\!\!,\,x\!\geqslant\! R_1,$$ the right side of which is evidently non-normalizable
at $+\infty$. Lemma~9 is proved.

{\bf Corollary 4.} In conditions of the Lemma 9 any formal associated function of $h$ of $n$-th order
normalizable at $\pm\infty$, for a spectral value $\lambda$ such that either $\lambda\leqslant0$ or ${\rm{Im}}\,
\lambda\ne0$, can be written in the form
\be\sum\limits_{j=0}^na_{j,\uparrow\downarrow}\varphi_{j,\uparrow\downarrow}(x),\qquad
a_{j,\uparrow\downarrow}={\rm{Const}},\quad a_{n,\uparrow\downarrow}\ne0\la{2.9} \ee and any  associated
function of $h$ of $n$-th order, non-normalizable at $\pm\infty$, for the same spectral value $\lambda$ can be
presented as follows \be\sum\limits_{j=0}^n\big(b_{j,\uparrow\downarrow}\varphi_{j,\uparrow\downarrow}(x)+
c_{j,\uparrow\downarrow}\hat\varphi_{j,\uparrow\downarrow}(x)\big),\ee where
$b_{j,\uparrow\downarrow},c_{j,\uparrow\downarrow}={\rm{Const}}$ and either $b_{n,\uparrow\downarrow}\ne0$ or
$c_{n,\uparrow\downarrow}\ne0.$

{\bf Corollary 5.} For normalizable associated functions $\psi_1(x)$ and $\psi_2(x)$ of a Hamiltonian $h\in K$
of any orders, for eigenvalues $\lambda_1$ and $\lambda_2$ respectively such that either $\lambda_{1,2}
\leqslant0$ or ${\rm{Im}}\,\lambda_{1,2}\ne0$ the equality \be\int\limits_{-\infty}^{+\infty}[h \psi_1(x)]
\psi_2(x)\,dx=\int\limits_{-\infty}^{+\infty} \psi_1(x) [h\psi_2(x)]\,dx\ee takes place.

\section{Invariance of the potential sets $\cal K$ and $K$}

Invariance of the potential sets $\cal K$ and $K$ under intertwining is proved in Lemmas 1 and 10 respectively.

{\bf Lemma 10.} {\it Let: 1) $h^+=-\partial^2+V_1(x)$, $V_1(x)\in\cal K$; 2) $\lambda\in\mathbb C$ and either
$\lambda\leqslant 0$ or ${\rm{Im}}\,\lambda\ne0$; 3) $\varphi(x)$ be zero-mode of $h^+-\lambda$; 4)
$\chi(x)=-\varphi'(x)/\varphi(x)$, $q_1^\pm=\mp\partial+\chi(x)$. Then the potential $V_2(x)$ of the Hamiltonian
$$h^-\equiv-\partial^2+V_2(x)=\lambda+q_1^-q_1^+,$$
intertwined  with $h^+=\lambda+q_1^+q_1^-$ by means of equalities
$$q_1^\pm h^\mp = h^\pm q_1^\pm$$
belongs to $\cal K$ also.}

{\bf Proof.}

Let us first check that there is $R'_{02}>0$ such that
$$V_2(x)\equiv V_1(x)-2(\ln\varphi(x))''$$
(see Eq. (53) in \cite{ACS}) for $|x|\geqslant R'_{02}$ is twice continuously differentiable.  For this purpose
it is sufficient to show  that there is $R'_{02}>0$ such that $\varphi(x)$ for $|x|\geqslant R'_{02}$ has not
zeroes and is four times continuously differentiable. Existence of $R'_{02}>0$ such that $\varphi(x)$ for
$|x|\geqslant R'_{02}$ has not zeroes follows from the fact that one of asymptotics of Lemma~8 is valid for
(normalized) $\varphi(x)$. Without loss of generality suppose that this $R'_{02}$ is so large that
\begin{equation}V_1(x)\Big|_{[R'_{02},+\infty[}\in
C^2_{[R'_{02},+\infty[},\qquad V_1(x)\Big|_{]-\infty,-R'_{02}]}\in
C^2_{]-\infty,-R'_{02}]}.\la{a27}\end{equation} Then the fact that $\varphi(x)$ is four times continuously
differentiable for $|x|\geqslant R'_{02}$ follows from the equality $\varphi'' = (V_1 - \lambda) \varphi$, from
\gl{a27} and from the fact that $\varphi(x)$  is twice continuously differentiable for $|x|\geqslant R'_{02}$ as
a zero-mode of $h^+-\lambda$.

Let us now verify that ${\rm{Im}\,}V_2/{\rm{Re}\,}V_2=o(1)$, $x\to\pm\infty$ and there are $R_{02}\geqslant
R'_{02}$ and $\varepsilon_2>0$ such that ${\rm{Re}\,}V_2(x)\geqslant\varepsilon_2$ for any $|x|\geqslant
R_{02}$. The former follows from \gl{a26} in view of ${\rm{Im}\,}V_1/{\rm{Re}\,}V_1 = o(1)$, $x\to\pm\infty$.
Moreover, since obviously
\begin{equation}
{\rm{Re}\,}V_2(x)={\rm{Re}\,}V_1(x)[1+o(1)],\qquad x\to\pm\infty\la{a28}\end{equation} and there are
$R_{02}\geqslant R'_{02}$ and $\varepsilon_1>0$ such that for any $|x|\geqslant R_{02}$, the value of $[1+o(1)]$
in \gl{a28} is more than or equal to $1/2$ and ${\rm{Re}\,}V_1(x)\geqslant\varepsilon_1$, so that for any
$|x|\geqslant R_{02}$ the inequalities hold
$${\rm{Re}\,}V_2(x)\geqslant{1\over2}{\rm{Re}\,}V_1(x)\geqslant{\varepsilon_1\over2},$$
wherefrom the existence of the required $R_{02}$ and $\varepsilon_2=\varepsilon_1/2$ follows.


Finally we show that the function
\begin{equation}\Big(\int\limits_{R_{02}}^x\sqrt{|V_2(x_1)|}\,dx_1\Big)^2
\Big({{|V'_2(x)|^2}\over{|V_2(x)|^3}}+{{|V''_2(x)|}\over{|V_2(x)|^2}}\Big) \la{a29}\end{equation} is bounded for
$x\geqslant R_{02}$ (the case with a similar function for $x\leqslant -R_{02}$ can be considered analogously).
In view of $V_1(x)\in\cal K$, \gl{a23}, \gl{a24} and \gl{a26} we have
$$\int\limits_{R_{02}}^x\sqrt{|V_2(x_1)|}\,dx_1=O\big(\xi_{1,\uparrow}(x)\big),
\qquad{{|V'_2(x)|^2}\over{|V_2(x)|^3}}=O\Big({1\over{\xi^2_{1,\uparrow}(x)}}\Big),$$
$${{|V''_2(x)|}\over{|V_2(x)|^2}}=O\Big({1\over{\xi^2_{1,\uparrow}(x)}}\Big),\qquad
x\to+\infty,\qquad\xi_{1,\uparrow}(x)=\int\limits_{R_{02}}^x\sqrt{|V_1(x)|}dx_1,$$ wherefrom boundedness of
\gl{a29} is derived. Lemma 10 is proved.

{\bf Corollary 6.} Using \gl{a13}, \gl{a15} and Remark~1, \gl{2.3} and \gl{2.5} (\gl{2.6} and \gl{2.8}),
\gl{a20}, \gl{2.9} and estimations similar to the estimations in the proof of Lemma~9, one can easily  check
that under conditions of Lemma~10 the operator $q_1^-$ maps any  formal eigenfunction or associated function (of
any order) of the Hamiltonian $h^+$ normalizable at $+\infty$ (at $-\infty$) to a  function normalizable at
$+\infty$ (at $-\infty$), for any spectral value $\lambda'$ such that either $\lambda'\leqslant0$ or
${\rm{Im}}\,\lambda'\ne0$.

{\bf Lemma 1.} {\it Let: 1) $h^+=-\partial^2+V_1(x)$, $V_1(x)\in K$; 2) $h^-=-\partial^2+V_2(x)$, $V_2(x)\in
C_{\mathbb R}$; 3) $q_N^-h^+=h^-q_N^-$, where $q_N^-$ is  a differential operator of $N$th order with
coefficients belonging to $C_{\mathbb R}^2$; 4) each eigenvalue of ${\bf S}^+$-matrix of $q_N^-$ (see Th. 1 in
Part~I \cite{ACS}) satisfies one of the conditions: either $\lambda\leqslant0$ or ${\rm Im}\, \lambda\ne0$.
Then: 1) $V_2(x)\in K$; 2) coefficients of $q_N^-$ belong to $C_{\mathbb R}^\infty$; 3) $h^+q_N^+=q_N^+h^-$,
where $q_N^+=(q_N^-)^t$, and moreover coefficients of $q_N^+$ belong to $C_{\mathbb R}^\infty$ as well.}

{\bf Proof.}

Let $\varphi_1(x)$, \dots, $\varphi_N(x)$ be a basis in ${\rm{ker}}\,q_N^-$, in which $\bf S^+$-matrix of
$q_N^-$ (see Theorem~1 in Part~I \cite{ACS}) has the canonical form. Since, firstly, $\varphi_1(x)$, \dots,
$\varphi_N(x)$ as eigen- and associated functions of $h^+$ belong to $C^\infty_{\mathbb R}$, secondly, the
Wronskian $W(x)$ of the functions $\varphi_1(x)$, \dots, $\varphi_N(x)$ has not any zeros  and, thirdly,
$$q_N^-={1\over{W(x)}}\begin{vmatrix}
\varphi_1(x) & \varphi'_1(x) & \dots & \varphi^{(N)}_1(x) \\
\varphi_2(x) & \varphi'_2(x) & \dots & \varphi^{(N)}_2(x) \\
\hdotsfor{4} \\
\varphi_N(x) & \varphi'_N(x) & \dots & \varphi^{(N)}_N(x) \\
1 & \partial & \dots & \partial^N
\end{vmatrix},$$
the coefficients of $q_N^-$ and thereby of $q_N^+$ belong to $C^\infty_{\mathbb R}$. Belonging of $V_2(x)$ to
$C^\infty_{\mathbb R}$ follows from the equality $V_2(x)=V_1(x)-2(\ln W(x))''$ (see Eq. (53) in \cite{ACS}),
from inclusion $W(x)\in C^\infty_{\mathbb R}$ and from absence of zeroes for $W(x)$. Inclusion $V_2(x)\in K$
follows from inclusion $V_2(x)\in C^\infty_{\mathbb R}$, from Lemma~10 and can be also justified by the
factorization procedure described in Lemma~1 of \cite{ansok}. The equality $h^+q_N^+=q_N^+h^-$ is obvious.
Lemma~1 is proved.


\section{Proofs of Lemmas 2--4 and Theorem 3}

The properties of associated functions under intertwining are described by the

{\bf Lemma 2.} {\it Let: 1) the conditions of the Lemma 1 take place; 2) $\varphi_n(x)$, $n=0$, \dots $M$ be a
sequence of formal associated functions of $h^+$ for spectral value $\lambda$:
$$h^+\varphi_0=\lambda\varphi_0,\qquad (h^+-\lambda)\varphi_n=
\varphi_{n-1},\quad n\geqslant1,$$ where either $\lambda\leqslant0$ or ${\rm{Im}}\,\lambda\ne0$. Then:

1) there is a number $m$ such that $0\leqslant m\leqslant\min\{M+1,N\}$,
$$ q_N^-\varphi_n\equiv0, \qquad n<m$$
and
$$\psi_l=q_N^-\varphi_{m+l},\qquad l=0,\ldots, M-m$$
is a sequence of formal associated functions of $h^-$ for the spectral value $\lambda$:
$$h^-\psi_0=\lambda\psi_0,\qquad (h^--\lambda)\psi_l=
\psi_{l-1},\quad l\geqslant1;$$

2) if a function $\varphi_n(x)$, for a given $0\leqslant n\leqslant M$, is normalizable at $+\infty$ (at
$-\infty$), then $q_N^-\varphi_n$ is normalizable at $+\infty$ (at $-\infty$) as well.}

{\bf Proof.}

Existence of $m$ such that $0\leqslant m\leqslant\min\{M+1,N\}$,
$$q_N^-\varphi_n\equiv0,\qquad n<m$$
and either $m>M$ or
\begin{equation}q_N^-\varphi_m\not\equiv0,\la{a51'''}\end{equation}
can be derived from linear independence of $\varphi_n$ and from the fact that dimension of ${\rm{ker}}\,q_N^-$
is $N$. The fact that $\psi_l=q_N^-\varphi_{m+l}$, $l=0$,~\dots, $M-m$ is a sequence of formal eigenfunction and
associated functions of $h^-$ (if $m\leqslant M$):
$$h^-\psi_0=\lambda\psi_0,\qquad (h^--\lambda)\psi_l=\psi_{l-1},\quad
l\geqslant1,$$ follows from the chains:
$$h^-\psi_0=h^-q_N^-\varphi_m=q_N^-h^+\varphi_{m}=q_N^-(\lambda
\varphi_{m}+\varphi_{m-1})=\lambda\psi_0,\qquad\varphi_{-1}\equiv0,$$
$$(h^--\lambda)\psi_l=(h^--\lambda)q_N^-\varphi_{m+l}=q_N^-(h^+-\lambda)
\varphi_{m+l}=q_N^-\varphi_{m+l-1}=\psi_{l-1},\qquad l\geqslant1,$$ if intertwining $h^-q_N^-=q_N^-h^+$ and
\gl{a51'''} are used. Before the proof of the second statement of the Lemma 2 let us note that with the help of
similar arguments one can show that in the conditions of Lemma~10 the operator $q_1^-$ maps any formal
eigenfunction or associated function of~$h^+$ for a spectral value $\lambda'$ either to the identical zero or to
a formal eigenfunction or associated function of~$h^-$ for the same spectral value $\lambda'$. Thus, the second
statement of the Lemma~2 follows from Lemma~10, corollary 6 and the construction, described in Lemma~1 of
\cite{ansok}. Lemma~2 is proved.

{\bf Corollary 7 (2).} Since $h^+$ is an intertwining operator to itself and both eigenvalues of its ${\bf
S^+}$-matrix (see Theorem~1 in Part~I \cite{ACS}) are zero, then if $\varphi_n(x)$ is normalizable at $+\infty$
(at $-\infty$), then $\varphi_j(x)$, $j=0$, \dots $n-1$ is normalizable at
$+\infty$ (at $-\infty$) as well.\\

{\bf Corollary 8 (3).} If there is a normalizable associated  function of $n$-th order $\varphi_n(x)$ of the
Hamiltonian $h$ with a potential belonging to $K$ for an eigenvalue $\lambda$, which is either
$\lambda\leqslant0$ or ${\rm {Im}}\,\lambda\ne0$, then for this eigenvalue there is an associated function
$\varphi_j(x)$ of the Hamiltonian $h$, normalizable on the whole axis, of any smaller order $j$:
$$\varphi_j=(h-\lambda)^{n-j}\varphi_n,\qquad j=0,\ldots,n-1.$$
\\

{\bf Corollary 9 (4).} Let $\varphi^-_{i,j}(x)$ be a canonical basis of zero-modes of the intertwining operator
$q_N^-$, {\it i.e.} such that ${\bf S^+} $-matrix (in Theorem 1 of Part~I \cite{ACS}) has in this basis the
canonical (Jordan) form:
$$h^+\varphi^-_{i,0}=\lambda_i\varphi^-_{i,0},\quad(h^+-\lambda_i)
\varphi^-_{i,j}=\varphi^-_{i,j-1},\quad i=1,\ldots, n,\quad j=1,\ldots, k_i-1,\qquad \sum\limits_{i=1}^nk_i=N.$$
Then there are numbers $k^+_{i\uparrow}$ and $k^+_{i\downarrow}$, $0\leqslant k^+_{i\uparrow,\downarrow}
\leqslant k_i$ such that for any $i$ the functions
$$\varphi^-_{i,j}(x),\qquad j=0,\ldots, k^+_{i\uparrow,\downarrow}-1$$
are normalizable at $+\infty$ or $-\infty$ respectively and the functions
$$\varphi^-_{i,j}(x),\qquad j=k^+_{i\uparrow,\downarrow},\ldots,k_i-1$$
are non-normalizable at the same $+\infty$ or $-\infty$.

Independence of these numbers $k^+_{i\uparrow,\downarrow}$ on a choice of the canonical basis in the case, when
the intertwining operator $q_N^-$ cannot be stripped-off,
follows from\\

{\bf Lemma 3.} {\it Let: 1) conditions of Lemma 1 take place; 2) $q_N^-$ not be able to be stripped-off. Then
any two formal associated functions of $h^+$ of the same order for the same spectral value $\lambda$ when being
zero-modes of $q_N^-$ are either simultaneously normalizable at $+\infty$ or simultaneously non-normalizable at
$+\infty$. The same takes place at $-\infty$.}

{\bf Proof.}

Assume that there are two sequences of a formal eigenfunction and associated functions of $h^+$ for the same
spectral value $\lambda$:
$$h^+\phi_{l,0}=\lambda\phi_{l,0},\qquad(h^+-\lambda)\phi_{l,j}=\phi_{l,j-1},
\quad l=1,2, \quad j=1,\ldots, j_0,$$ such that $\phi_{1,j_0}$ is normalizable at $+\infty$, $\phi_{2,j_0}$ is
non-normalizable at $+\infty$ and
$$q_N^-\phi_{l,j_0}=0,\qquad l=1,2.$$
Let us show that it leads to contradiction.

Let us check first that
$$q_N^-\phi_{l,j}=0,\qquad l=1,2, \quad j=0,\ldots, j_0-1.$$
For $j=j_0-1$ these equalities follow from the chain
$$q_N^-\phi_{l,j_0-1}=q_N^-(h^+-\lambda)\phi_{l,j_0}=(h^--\lambda)q_N^-
\phi_{l,j_0}=0,$$ and for $j<j_0-1$ they can be derived in the same way by induction. As for intertwining
operator, which cannot be stripped-off, there is only one zero-mode of $h^+-\lambda$ (up to a constant
cofactor), corresponding to a fixed eigenvalue $\lambda$ of its $\bf S^+$-matrix (see Th.~1\&2 in Part~I
\cite{ACS}), so $\phi_{1,0}(x)$ and $\phi_{2,0}(x)$ are proportional. Without loss of generality suppose that
$\phi_{1,j}$ and $\phi_{2,j}$ are normalized so that
$$\phi_{1,0}(x)\equiv\phi_{2,0}(x).$$
Then the sequence $\phi_{1,j}-\phi_{2,j}$ represents a sequence of associated functions  of $h^+$ for the same
eigenvalue $\lambda$ (being zero-modes of $q_N^-$) and $\phi_{1,j_0}-\phi_{2,j_0}$ is an associated function of
the order $j_1<j_0$ non-normalizable at $+\infty$. But on the other hand there is
 an associated
function $\phi_{1,j_1}$ of $h^+$  normalizable at $+\infty$ (see corollary~7 (2)) of the order $j_1$ for an
eigenvalue $\lambda$ which is a zero-mode of $q_N^-$. Performing in the same way by induction, we come to the
conclusion that intersection ${\rm{ker}}\,q_N^-\cap{\rm{ker}}\,(h^+-\lambda)$ (dimension of which  is~1 in view
of Th.~2 of Part~I \cite{ACS}) contains non-trivial  functions normalizable and non-normalizable at~$+\infty$,
the latter being impossible. The consideration of the $-\infty$ case is analogous. The Lemma~3 is proved.

The following Lemma 4 clarifies interrelation between the behavior at $\pm\infty$ of elements of canonical bases
of mutually transposed intertwining operators.

{\bf Lemma 4.} {\it Let: 1) conditions of Lemma 1 take place; 2) $\{\varphi^-_{i,j}\}$ and $\{\varphi^+_{i,j}\}$
are canonical bases of ${\rm{ker}}\,q_N^-$ and ${\rm{ker}}\,q_N^+$ respectively; 3) $q_N^-$ cannot be
stripped-off; 4) $k_i$ is algebraic multiplicity of eigenvalue $\lambda_i$ of $\bf S^+$-matrix (see Th. 1 of
Part~I \cite{ACS}). Then for any $i$ and $j$ the function $\varphi_{i,j}(x)$ is normalizable (non-normalizable)
at $+\infty$ if and only if $\psi_{i, k_i-j-1}(x)$ is non-normalizable (normalizable) at $+\infty$. The same
takes place at $-\infty$.}

{\bf Proof.}

In accordance with corollary 9 (4) for any $i$ the basis $\varphi^-_{i,j}$ has  the following structure:
$$\varphi^-_{i,j},\qquad j=0,\ldots,k^+_{i\uparrow,\downarrow}-1$$
are normalizable at $\pm\infty$ and
$$\varphi^-_{i,j},\qquad j=k^+_{i\uparrow,\downarrow},\ldots,k_{i}-1$$
are non-normalizable at $\pm\infty$, where $0\leqslant k^+_{i\uparrow,\downarrow} \leqslant k_i$. Moreover in
view of Lemma~3 the numbers $k^+_{i\uparrow,\downarrow}$ are independent of a choice of a canonical basis. To
prove Lemma~4 it is sufficient to establish that for any $i$
\begin{equation}\varphi^+_{i,j},\qquad j=0,\ldots,k_i-k^+_{i\uparrow,\downarrow}-1\la{a52}
\end{equation}
are normalizable at $\pm\infty$ and
\begin{equation}\varphi^+_{i,j},\qquad j=k_i-k^+_{i\uparrow,\downarrow},\ldots,k_{i}-1\la{a53}
\end{equation}
are non-normalizable at $\pm\infty$.

Let $\varphi_{i,j,\uparrow\downarrow}$ be a sequence of a formal eigenfunction and associated functions of $h^+$
normalizable at $\pm\infty$, for a spectral value $\lambda_i$. Then because of Lemma~3
$$q_N^-\varphi_{i,j,\uparrow\downarrow}\not\equiv0,\qquad j=k^+_{i\uparrow,
\downarrow},\ldots,k_i-1.$$ On the other hand, by virtue of Lemma~2,  the functions
$q_N^-\varphi_{i,j,\uparrow\downarrow}$ form a sequence of a formal eigenfunction and associated functions of
$h^-$ for the same spectral value $\lambda_i$ and for $j\leqslant k_i -1$ represent  zero-modes of $q_N^+$
(since in virtue of Th.~1 of Part~I \cite{ACS}, $q_N^+q_N^-$  is a polynomial of $h^+$, containing
cofactor~$(h^+-\lambda_i)^{k_i}$). Moreover in view of Lemma~2 these functions  are normalizable at $\pm\infty$.
Thus, by virtue of Lemma~3 the functions \gl{a52} are normalizable at $\pm\infty$.

We prove now that the functions \gl{a53} are non-normalizable at $\pm\infty$. For this purpose, because of
corollary~9 (4), it is sufficient to prove that $\varphi^+_{i,k_i-k^+_{i\uparrow,\downarrow}}$ is
non-normalizable at $\pm\infty$. Let us consider factorization of $q_N^-$ in the product of intertwining
operators of first order in accordance with Lemma~1 of \cite{ansok}:
$$q_N^-=r_N^-\,\ldots\, r_1^-,$$
where $r_1^-$, \dots, $r_{k_i}^-$ are chosen so that
$$r_{j+1}^-\,\ldots\, r_1^-\varphi^-_{i,j}=0,\qquad j=0,\ldots,k_i-1.$$
Then for $q_N^+$ there is a factorization
$$q_N^+=(r_1^-)^t\,\ldots\,(r_N^-)^t,$$
where the zero-mode of $(r_j^-)^t$ is evidently
$$(r_{j+1}^-)^t\,\ldots\,(r_N^-)^t\varphi^+_{i,k_i-j},\qquad j=1,\ldots,k_i.$$
Suppose that $\varphi^+_{i,k_i-k^+_{i\uparrow,\downarrow}}$ is normalizable at $\pm\infty$. Then
$$(r_{k^+_{i,\uparrow\downarrow}+1}^-)^t\,\ldots\,(r_N^-)^t\varphi^+_{i,k_i-k^+_{i\uparrow,\downarrow}}$$
({\it i.e.} in view of corollary~6 the zero-mode of $(r_{k^+_{i\uparrow,\downarrow}}^-)^t$)  is normalizable at
$\pm\infty$ as well. But this statement contradicts to the fact that
$$r_{k^+_{i\uparrow,\downarrow}-1}^-\,\ldots\,r_1^-\varphi^-_{i,k^+_{i\uparrow,\downarrow}-1}$$
({\it i.e.} the zero-mode of $r_{k^+_{i\uparrow,\downarrow}}^-$) is normalizable (because of the same
corollary~6) at $\pm\infty$. Thus, $\varphi^+_{i,k_i-k^+_{i\uparrow,\downarrow}}$ is non-normalizable at
$\pm\infty$ and Lemma~4 is proved.

A more precise result on interrelation between Jordan structures of intertwined Hamiltonians and the behavior of
transformation functions is contained in

{\bf Theorem 3.} {\it Let: 1) the  conditions of Lemma 4 take place; 2) $\nu_\pm(\lambda)$ is an algebraic
multiplicity of an eigenvalue $\lambda$ of $h^\pm$, {\it i.e.} the number of independent eigenfunctions and
associated
functions of $h^\pm$ 
normalizable on the whole axis; 3) if $\lambda$ is not an eigenvalue of $\bf S^+$ (see Th. 1 of Part~I
\cite{ACS}), then $n_+(\lambda)=n_-(\lambda)=n_0(\lambda)=0$ and if $\lambda= \lambda_i$, where $\lambda_i$ is
an eigenvalue of $\bf S^+$, then $n_\pm(\lambda_i)$ is a number of  functions among $\varphi^\mp_{i,j}(x)$,
$j=0$, \dots, $k_i-1$ normalizable at both infinities and $n_0(\lambda_i)$ is a number of functions among
$\varphi_{i,j}^-(x)$ (or $\varphi_{i,j}^+(x)$), $j=0$, \dots, $k_i-1$ normalizable only at one of infinities.
Then for any $\lambda$ such that either $\lambda\leqslant0$ or ${\rm {Im}}\,\lambda\ne0$ the equality
$$\nu_+(\lambda)-n_+(\lambda)=\nu_-(\lambda)-n_-(\lambda)$$
takes place. Moreover if $n_0(\lambda)>0$ for some $\lambda$, then for this $\lambda$
$$\nu_+(\lambda)-n_+(\lambda)=\nu_-(\lambda)-n_-(\lambda)=0.$$
}

{\bf Proof.}
Let us first notice that if for the level of the Hamiltonian $h^+$ $\lambda$ such that either
$\lambda\leqslant0$ or ${\rm{Im}\,}\lambda\ne0$ there is an associated function of the $l$-th order normalizable
on the whole axis, then for the same level $\lambda$, any  associated function of $h^+$ of the $l$-th order
normalizable at one of infinities is normalizable on the whole axis. This fact is easily verifiable in a way
similar to the proof of Lemma~3. Thus, in the case $n_0(\lambda)>0$ there is no any
 associated function   of $h^+$ normalizable on the whole axis, of the order
$n_+(\lambda)$ (and consequently of any greater order) for the level $\lambda$. Hence in this case
$\nu_+(\lambda)=n_+(\lambda)$. Moreover in view of Lemma~4 and of the symmetry between $h^+$ and $h^-$ the
equality $\nu_- (\lambda)=n_-(\lambda)$ holds  for the case $n_0(\lambda)>0$ as well. Thus, for this case
Theorem~3 is proved.

In the case, when $\lambda$ does not belong to the spectrum of $\bf S^+$-matrix (see Th.~1 in Part~I \cite{ACS})
Theorem~3 follows from Lemma~2 and from the fact that an associated function of $h^\pm$, which corresponds to
$\lambda$ under consideration, cannot be zero-mode of $q_N^\mp$ (since in the opposite case this function would
be linear combination of formal eigenfunctions and associated functions of $h^\pm$, whose eigenvalues belong to
the spectrum of $\bf S^+$).

Consider now the case, when $\lambda=\lambda_i$ belongs to the spectrum of $\bf S^+$ and $n_0(\lambda_i)=0$. We
shall prove the inequality
\begin{equation}\nu_-(\lambda_i)-\nu_+(\lambda_i)+n_+(\lambda_i)\leqslant
n_-(\lambda_i) \la{a54}\end{equation} only, because the opposite inequality
$$\nu_+(\lambda_i)-\nu_-(\lambda_i)+n_-(\lambda_i)\leqslant n_+(\lambda_i)$$
 follows
from \gl{a54}, Lemma~4 and the symmetry between $h^+$ and $h^-$ (the statement of the theorem  is derived from
these inequalities). Since in the subcase $\nu_-(\lambda_i)-\nu_+(\lambda_i)+n_+(\lambda_i)\leqslant0$ the
inequality \gl{a54} is trivial,  we shall consider below the subcase
\begin{equation}\nu_-(\lambda_i)-\nu_+(\lambda_i)+n_+(\lambda_i)>0
\la{a55}\end{equation} only.

Let us show that there is a sequence $\hat\varphi_{i,j}(x)$ such that,
$$h^+\hat\varphi_{i,0}=\lambda_i\hat\varphi_{i,0},\qquad
(h^+-\lambda_i)\hat\varphi_{i,j}=\hat\varphi_{i,j-1},\quad j=1,\dots, \nu_-(\lambda_i)+n_+(\lambda_i)-1$$
$$q_N^-\hat\varphi_{i,j}=0,\qquad j=0,\ldots, \nu_-(\lambda_i)+n_+(\lambda_i)-1$$
and the functions $\hat\varphi_{i,j}$, $j=\nu_+(\lambda_i)$, \dots, $\nu_-(\lambda_i)+n_+(\lambda_i)-1$ are
non-normalizable at both infinities. This sequence cannot contain more than $k_i$ terms, since in the opposite
case associated functions of this sequence of orders greater $k_i-1$ would be linear combinations of
$\varphi^-_{i,j}$, the latter being impossible. Therefore, in view of Lemma~3,  the number of
 associated functions of the
sequence non-normalizable at both infinities cannot be greater than the number of functions non-normalizable at
both infinities among $\varphi_{i,j}^-$ with fixed $i$,
$$\nu_-(\lambda_i)-\nu_+(\lambda_i)+n_+(\lambda_i)\leqslant n_-(\lambda_i),$$
that is required to be proved.

Consider a sequence of  $\varphi_{i,j,\uparrow\downarrow}$  formal eigenfunction and associated functions of
$h^+$ normalizable at $\pm\infty$, for the level $\lambda_i$ (this sequence exists due of Lemma~9). First
$\nu_+(\lambda_i)$ functions of this sequence are normalizable at both infinities  (following the arguments used
at the  beginning of this proof). By virtue of \gl{2.9} any of the functions $\varphi^-_{i,j}$, $j=0,$~\dots,
$n_+(\lambda_i)-1$  can be presented as a linear combination of $\varphi_{i,0,\uparrow\downarrow}$,~\dots,
$\varphi_{i,n_+(\lambda_i)-1,\uparrow\downarrow}$. Moreover, due to linear independence of
$\varphi^-_{i,0}$,~\dots, $\varphi^-_{i,n_+(\lambda_i)-1}$ the reverse is valid as well. Hence,
$$q_N^-\varphi_{i,j,\uparrow\downarrow}=0,\qquad j=0,\ldots,n_+(\lambda_i)-1.$$
Moreover, in view of Lemmas~2 and~3 the functions
$$q_N^-\varphi_{i,j,\uparrow\downarrow},\qquad j=n_+(\lambda_i),\ldots$$
are different from  zero and form a sequence of a  formal eigenfunction and associated functions of $h^-$
normalizable at $\pm\infty$ for the level $\lambda_i$. Applying the arguments of the beginning of this proof one
can show that the  first $\nu_-(\lambda_i)$ terms of this sequence are normalizable at both infinities.

Using the sequence of formal associated functions~$\varphi_{i,j,\uparrow\downarrow}$ one can construct another
sequence of formal associated functions of $h^+$ for the same level~$\lambda_i$,
$$\tilde\varphi_{i,j,\uparrow\downarrow}=\sum_{k=0}^jA_{i,k,\uparrow\downarrow}
\varphi_{i,j-k,\uparrow\downarrow}, \qquad A_{i,k,\uparrow\downarrow}={\rm{Const}},\quad
A_{i,0,\uparrow\downarrow}\ne0.$$ This sequence  as well as the sequence $\varphi_{i,j,\uparrow\downarrow}$ has
the following properties:
\begin{itemize}
\item $\tilde\varphi_{i,j,\uparrow\downarrow}$, $j=0$, \dots \ are normalizable at $\pm\infty$;
\item $\tilde\varphi_{i,j,\uparrow\downarrow}$, $j=0$, \dots, $\nu_+(\lambda_i)-1$ are
normalizable at both infinities;
\item
\begin{equation}q_N^-\tilde\varphi_{i,j,\uparrow\downarrow}=0,\qquad
j=0,\ldots,n_+(\lambda_i)-1;\la{a56}\end{equation}
\item
$q_N^-\tilde\varphi_{i,j,\uparrow\downarrow},\qquad j=n_+(\lambda_i),\ldots$ are different from zero and form
the sequence of  formal eigenfunction and associated functions of $h^-$ normalizable at $\pm\infty$ for the
level $\lambda_i$;
\item
$q_N^-\tilde\varphi_{i,j,\uparrow\downarrow},\qquad j=n_+(\lambda_i),\ldots,\nu_-(\lambda_i)+n_+(\lambda_i)-1$
are normalizable at both infinities.
\end{itemize}
One can choose constants $A_{i,k,\uparrow\downarrow}$ so that the required sequence $\hat\varphi_{i,j}$ can be
written in the form
$$\hat\varphi_{i,j}=\tilde\varphi_{i,j,\uparrow}-\tilde\varphi_{i,j,\downarrow}.$$

Indeed, notice that from inequalities \gl{a55} and $\nu_+(\lambda_i)\geqslant n_+(\lambda_i)$ it follows that
$\nu_-(\lambda_i)>0$, {\it i.e.} that there is a normalizable eigenfunction $\psi_{i,0}$ of $h^-$ for the level
$\lambda_i$. As there is only one (up to constant cofactor) normalizable eigenfunction of $h^-$ for the level
$\lambda_i$
 the equalities
$$q_N^-\varphi_{i,n_+(\lambda_i),\uparrow\downarrow}=C_{i,\uparrow\downarrow}\psi_{i,0}$$ take  place for some constants $C_{i,\uparrow\downarrow}\ne0$.
The fact that the relation
\begin{equation}q_N^-\hat\varphi_{i,j}=0\la{a57}\end{equation}
holds for $j=0$, \dots, $n_+(\lambda_i)-1$ follows from \gl{a56}. The equality \gl{a57} holds for
$j=n_+(\lambda_i)$ if we take $A_{i,0,\uparrow\downarrow}= C_{i,\downarrow\uparrow}$ since
$$q_N^-\hat\varphi_{i,n_+(\lambda_i)}=\sum\limits_{k=0}^{n_+(\lambda_i)}
\Big(A_{i,k,\uparrow}q_N^-\varphi_{i,n_+(\lambda_i)-k,\uparrow}-A_{i,k,\downarrow}
q_N^-\varphi_{i,n_+(\lambda_i)-k,\downarrow}\Big)=$$
$$A_{i,0,\uparrow}q_N^-\varphi_{i,n_+(\lambda_i),\uparrow}-A_{i,0,\downarrow}
q_N^-\varphi_{i,n_+(\lambda_i),\downarrow}=
\big(A_{i,0,\uparrow}C_{i,\uparrow}-A_{i,0,\downarrow}C_{i,\downarrow}\big)\psi_{i,0}=0.$$ At last, one can
attain validity of \gl{a57} for $j=n_+(\lambda_i)+1$,~\dots, $\nu_-(\lambda_i)+n_+(\lambda_i)-1$, looking
through all $j=n_+(\lambda_i)+1$,~\dots, $\nu_-(\lambda_i)+n_+(\lambda_i)-1$ and taking into account at every
step that $q_N^-\hat\varphi_{i,j}$ being normalizable eigenfunction of $h^-$ is proportional to $\psi_{i,0}$.
One has also to take into account that the dependence $q_N^-\hat\varphi_{i,j}$ of
$A_{i,j-n_+(\lambda_i),\uparrow\downarrow}$ is linear,
$$q_N^-\hat\varphi_{i,j}=\sum\limits_{k=0}^j\Big(A_{i,k,\uparrow}q_N^-\varphi_{i,j-k,\uparrow}
-A_{i,k,\downarrow}q_N^-\varphi_{i,j-k,\downarrow}\Big)$$
$$=\sum\limits_{k=0}^{j-n_+(\lambda_i)}\Big(A_{i,k,\uparrow}q_N^-\varphi_{i,j-k,\uparrow}
-A_{i,k,\downarrow}q_N^-\varphi_{i,j-k,\downarrow}\Big)$$ $$=\sum\limits_{k=0}^{j-n_+(\lambda_i)-1}
\Big(A_{i,k,\uparrow}q_N^-\varphi_{i,j-k,\uparrow}-A_{i,k,\downarrow}
q_N^-\varphi_{i,j-k,\downarrow}\Big)+\big(A_{i,j-n_+(\lambda_i),\uparrow}C_{i,\uparrow}-A_{i,j-n_+(\lambda_i),\downarrow}
C_{i,\downarrow}\big)\psi_{i,0},$$ and choose $A_{i,j-n_+(\lambda_i),\uparrow\downarrow}$ so that the
proportionality coefficient between $q_N^-\hat\varphi_{i,j}$ and $\psi_{i,0}$ is vanishing. It happens that
among $\hat\varphi_{i,j}$ there are $\nu_-(\lambda_i) +n_+(\lambda_i)-\nu_+(\lambda_i)$ functions, which are
non-normalizable at both infinities, because $\tilde\varphi_{i,j,\uparrow\downarrow}$,
$j=\nu_+(\lambda_i)$,~\dots, $\nu_-(\lambda_i)+n_+(\lambda_i)-1$ are normalizable at $\pm\infty$ only. Thus, the
required sequence is constructed and Theorem~3 is proved.
\section{Acknowledgements}
I am grateful to my collaborators A.A.Andrianov and F.Cannata for invaluable remarks, fruitful discussions and
reading the manuscript. This work  was supported by Grant RFBR 06-01-00186-a and partially supported by the INFN
grant.

\end{document}